\documentclass[conference]{IEEEtran}

\usepackage{cite}
\ifCLASSINFOpdf
  \usepackage[pdftex]{graphicx}
    \graphicspath{{.}}
      \DeclareGraphicsExtensions{.pdf,.jpeg,.png}
\else
                  \fi

\usepackage[cmex10]{amsmath}

\usepackage{csvsimple}
\usepackage{color} \usepackage[usenames,dvipsnames,svgnames]{xcolor}
\usepackage{microtype}
\usepackage{vwcol}  
\usepackage{multirow}
\usepackage{multicol}
\usepackage{pifont}
\usepackage[T1]{fontenc}
\usepackage{subcaption}
\usepackage{textgreek}

\newcommand{\code}[1]{\mbox{\texttt{#1}}}

\usepackage{listings}
\lstdefinestyle{llvm}{
 morekeywords={phi, add, icmp, eq, br, label, sext, to},
 emph={i64, i32, i1},
 emphstyle={\color{darkgray}\textbf},
}

\mathchardef\hyphenmathcode=\mathcode`\-

\let\origlstlisting=\lstlisting
\let\endoriglstlisting=\endlstlisting
\renewenvironment{lstlisting}
    {\mathcode`\-=\hyphenmathcode
     \everymath{}\mathsurround=0pt\origlstlisting}
    {\endoriglstlisting}

\usepackage{array}
\usepackage{mdwmath}
\usepackage{mdwtab}
\usepackage{float}
\newfloat{program}{t}{lop}
\usepackage{fixltx2e}
\usepackage[hidelinks]{hyperref}
\usepackage{url}
\usepackage{xcolor}
\usepackage{xspace}

\usepackage[comma,square,sort&compress]{natbib}
\newcommand{\myparagraph}[1]{\noindent \textbf{#1}}

\newcommand{\secref}[1]{$\S$\ref{sec:#1}}

\newcommand{\figref}[1]{Figure~\ref{fig:#1}}
\newcommand{\figreftwo}[2]{Figures~\ref{fig:#1} and~\ref{fig:#2}}

\newcommand{\tabref}[1]{Table~\ref{tab:#1}}
\newcommand{\tabreftwo}[2]{Tables~\ref{tab:#1} and~\ref{tab:#2}}

\newcommand{\projectname}{\textsc{Elzar}\xspace}
\newcommand{\swiftr}{\textsc{SWIFT-R}\xspace}
\newcommand{\intelavx}{Intel AVX\xspace}
\newcommand{\intel}{Intel\xspace}
\newcommand{\ilr}{\mbox{ILR}}

\newcommand{\flagsreg}{\mbox{FLAGS}}

\let\labelindent\relax
\usepackage[inline]{enumitem}
\setlist{noitemsep,topsep=0pt,parsep=0pt,partopsep=0pt}

\usepackage{setspace}
\usepackage[margin=5pt,font={stretch=0.9}]{caption}

\begin{document}
\title{\huge \projectname: Triple Modular Redundancy using \\ \intel Advanced Vector Extensions\\[3mm] \Large Technical Report}
\author{	\IEEEauthorblockN{Dmitrii Kuvaiskii$^\dagger$ \quad Oleksii Oleksenko$^\dagger$ \quad Pramod Bhatotia$^\dagger$ \quad Pascal Felber$^\ddagger$ \quad Christof Fetzer$^\dagger$}
	\IEEEauthorblockA{$^\dagger$ Technical University of Dresden, Germany\\
		$^\ddagger$ University of Neuch\^atel, Switzerland}}

\maketitle
\thispagestyle{plain}
\pagestyle{plain}

\begin{abstract}
Instruction-Level Redundancy (\ilr) is a well known approach to tolerate transient CPU faults.
It replicates instructions in a program and inserts periodic checks to detect and correct CPU faults using majority voting, which essentially requires three copies of each instruction and leads to high performance overheads.
As SIMD technology can operate simultaneously on several copies of the data, it appears to be a good candidate for decreasing these overheads.
To verify this hypothesis, we propose \projectname{}, a compiler framework that transforms unmodified multithreaded applications to support triple modular redundancy using \intelavx extensions for vectorization.
Our experience with several benchmark suites and real-world case-studies yields mixed results: while SIMD may be beneficial for some workloads, e.g., CPU-intensive ones with many floating-point operations, it exhibits higher overhead than \ilr{} in many applications we tested.
We study the sources of overheads and discuss possible improvements to \intelavx that would lead to better performance. 
\end{abstract}
 
\IEEEpeerreviewmaketitle

\section{Introduction}
\label{sec:intro}

Transient faults in CPUs can cause arbitrary state corruption during computation. Therefore, they pose a significant challenge for software systems reliability \cite{CPUFailures2005}.
The causes for transient faults are manifold, including radiation/particle strikes, dynamic voltage scaling, manufacturing variability, device aging, etc. \cite{Borkar2005}.
Moreover, the general trend of ever-decreasing transistor sizes with lower operating voltages only worsens the reliability problem \cite{Henkel2013,DarkSilicon2014}.

The unreliability of CPUs is especially threatening at the scale of data centers, where tens of thousands of machines are used to support modern online services.
At this sheer scale, CPU faults happen at a surprisingly high rate and tend to increase in frequency after the first occurrence, as reported by a number of large-scale in-the-field studies \cite{schroeder2010large,CyclesCells2011,BugsInCloud2014}. Since the machines in data centers operate in tight collaboration, a single CPU fault can propagate to the entire data center, leading to catastrophic consequences \cite{AmazonS32008,AmazonLoadBalancer2008}.

To overcome the problem of transient CPU faults, large-scale online services started using ad-hoc mechanisms such as integrity checks, checksums, etc. For instance, Mesa \cite{Mesa2014}, a data warehousing system at Google, makes use of application-specific integrity checks to detect transient faults during computation.
Unfortunately, ad-hoc mechanisms have two major limitations: (1)~they require manual effort to design and implement application-specific integrity checks, and (2)~they can only protect from errors that are anticipated by the application programmer.

As an alternative to ad-hoc checking techniques, one can make use of a principled approach like Byzantine Fault Tolerance (BFT). BFT-based systems do not only tolerate transient faults, but also malicious adversaries. Unfortunately, BFT yields high performance and management overheads because of its broad assumptions on the type of faults and the power of the adversary \cite{BFTSceptics2009, Bhatotia2010}. Since most online services run behind the security perimeter of a data center, the ``pessimistic'' BFT fault model is considered overkill.  Therefore, BFT-based systems are rarely adopted in practice.

To find a good compromise between ad-hoc mechanisms and BFT-based systems, a number of light-weight \emph{hardening} techniques were proposed (see $\S$\ref{sec:rel}). These hardening techniques transform the original program to locally detect and correct faults. A well-known hardening approach is Instruction-Level Redundancy (\ilr) \cite{EDDI2002,Swift2005,haft2016}. \ilr{} is a compile-time transformation that replicates original instructions to create separate data flows and inserts periodic checks to detect divergence caused by transient faults in these data flows. In particular, \ilr{} duplicates instructions to achieve fault detection \cite{EDDI2002,Swift2005} and triplicates them to tolerate faults by majority voting \cite{SwiftR2007}.

\begin{figure}[t]
\centering
\includegraphics[scale=0.7, angle=-90]{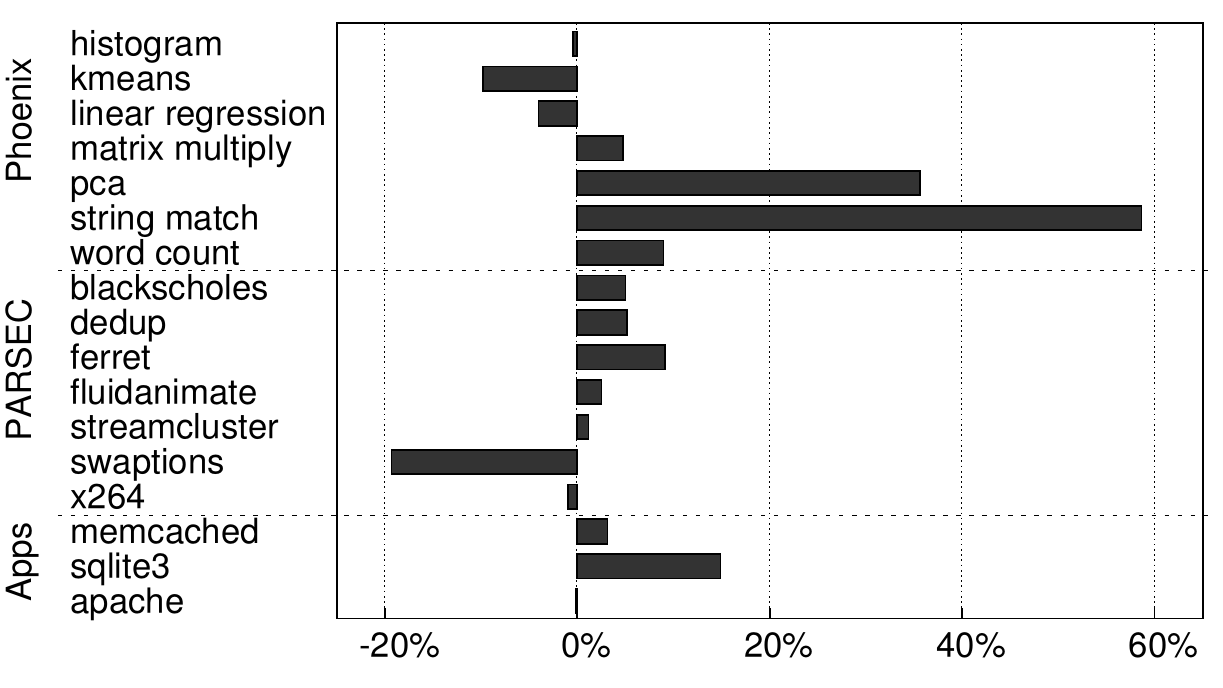}
\caption{Performance improvement with SIMD vectorization enabled (maximum runtime speedup for Phoenix and PARSEC benchmarks, maximum throughput increase for Memcached, SQLite3, and Apache).
}\label{fig:benches-vectorization-impact}
\vspace{2mm}
\end{figure}
 
As a result, with \ilr{} the CPU executes \emph{the same instruction} two or three times on \emph{several data copies}. We notice that, in fact, this corresponds to the very definition of Single Instruction Multiple Data (SIMD) processing. SIMD exploits data level parallelism, i.e., a single instruction operates on several pieces of data in parallel. Given that most modern CPUs have support for SIMD processing (Intel x86's SSE and AVX, IBM Power's AltiVec, and ARM's Neon), we
can naturally ask the following question: \emph{Can we utilize SIMD instructions to tolerate transient CPU faults and 
achieve better performance than {\ilr} with three copies?}

Before answering this question, we first need to understand how much of the SIMD potential of modern CPUs is {\em actually} being used in real-world applications. To investigate this, we tested applications from the Phoenix \cite{Phoenix2007} and PARSEC \cite{Parsec2009} benchmark suites, as well as several real-world applications, namely Memcached, SQLite, and the Apache web server. We compiled all applications in two versions: ``native'' with all optimizations enabled, and ``no-SIMD'' where we disable SSE, AVX, and all vectorization optimizations in LLVM. The performance improvements of native over no-SIMD, shown in \figref{benches-vectorization-impact}, indicate that most applications do not utilize the benefits of SIMD processing. Indeed, most of them exhibit less than $10\%$ improvement, with only \emph{string match} significantly benefiting from AVX.\footnote{Some applications (e.g., \emph{kmeans} and \emph{swaptions}) actually perform worse when SIMD is enabled. This counter-intuitive result is explained by the fact that compilers have only rough cycle-cost models and sometimes produce suboptimal instruction sequences.}
One can therefore conclude that SIMD processing units are currently largely underutilized CPU resources and could hence be used for fault tolerance.

To this end, we propose \projectname{},\footnote{Named after a four-armed character of Futurama. Similarly, \intelavx has $4\times64$-bit wide registers for SIMD processing.} a  compiler framework to harden \emph{unmodified multithreaded} programs by leveraging SIMD instructions available in modern CPUs (\secref{design}). \projectname{} is built on the \intelavx technology to achieve triple modular redundancy.
Since AVX possesses  256-bit wide registers and regular programs operate on at most 64-bit ones, it is possible to operate with four replicas in parallel, which is more than enough to harden applications and mask faults with majority voting.
Consequently, if a hardware fault affects one of the four replicas in an AVX register, it can be detected and outvoted by the other, correct replicas.

We implemented \projectname{} as an extension of the LLVM compiler framework (\secref{impl}).
It executes as a pass of the usual build process right before the final code generation.
In particular, \projectname{} transforms all the regular instructions of an application into their AVX-based counterparts, replicating data throughout AVX registers.
To achieve such transparent transformation, we use a mix of LLVM vectors and low-level AVX intrinsics.

We evaluated our approach by applying \projectname{} to the Phoenix and PARSEC benchmark suites (\secref{eval}), as well as three real-world case-studies: Memcached, SQLite3, and Apache (\secref{studies}).
To our disappointment, our evaluation showed mostly negative results, with an average normalized runtime slowdown of $4.1$--$5.6\times$ depending on the number of threads.
When compared against a straightforward instruction triplication approach \cite{SwiftR2007}, \projectname{} performed $46\%$ worse on average.
At the same time, \projectname{} was better on CPU-intensive benchmarks with few memory accesses and many floating-point operations.

We attribute poor performance of \projectname{} to two main causes.
First, there is a significant discrepancy between the regular CPU instructions and their AVX counterparts.
This discrepancy forced us to introduce additional wrapper instructions that significantly hamper performance.
Second, AVX instructions in general have higher latencies and are less optimized than the regular CPU instructions.
Nonetheless, we believe there is potential in using AVX for fault tolerance, and discuss how future implementations of this technology could boost \projectname's performance via minor modifications to the AVX instruction set (\secref{disc}).
Our rough estimation suggests that \projectname{} could achieve overheads as low as $48\%$ with the changes we propose.

 \section{Background and Related Work}
\label{sec:rel}

Our approach is based on three ideas: software-based hardening for fault detection, triple modular redundancy for fault recovery, and \intelavx{} technology for SIMD-based fault tolerance.

\subsection{Software-Based Hardening}
\label{sec:software_redundancy}

Software-based hardening techniques can be broadly divided into three categories: Thread-Level Redundancy (TLR) also called Redundant Multithreading (RMT), Process-Level Redundancy (PLR), and Instruction-Level Redundancy (\ilr).
	
\myparagraph{Redundant Multithreading (RMT).}
In RMT approaches \cite{HWRMT2002,DAFT2010}, a hardened program spawns an additional \emph{trailing} thread for each original thread. At runtime, trailing threads are executed on separate spare cores or take advantage of the Simultaneous Multithreading (SMT) capabilities of modern CPUs. Similar to \projectname, RMT allows keeping only one memory state among replicas (assuming that memory is protected via ECC). However, RMT approaches heavily rely on the assumption of spare cores or unused SMT, which is commonly not the case in multithreaded environments where programs tend to use all available CPU cores.

\myparagraph{Process Level Redundancy (PLR).}
PLR implements the similar idea as RMT, but at the level of separate processes \cite{PLR2007,RAFT2012}. In PLR, each process replica operates on its own memory state, and all processes synchronize on system calls. In multithreaded environments, allocating a separate memory state for each process raises a challenge of non-determinism because memory interleavings can result in discrepancies among processes and lead to false positives. Some PLR approaches resolve this challenge by enforcing deterministic multithreading \cite{RomainMT2014}. PLR might incur a lower performance overhead than RMT but it still requires spare cores for efficient execution.

\myparagraph{Instruction-Level Redundancy (\ilr).}
In contrast to RMT and PLR, ILR performs replication \emph{inside} each thread and does not require additional CPU cores \cite{EDDI2002, Swift2005}.
This in-thread replication seamlessly enables multithreading and requires no spare cores for performance.
We present \ilr{} in detail in \secref{ilr}.

Recent work on \ilr{} mainly concentrated on optimizations to trade-off fault coverage for lower overheads \cite{ESoftCheck2009, Shoestring2010}. In contrast to these new approaches, \projectname{} aims to utilize SIMD technology available on modern CPUs to achieve low performance overhead without compromising on fault coverage. 
A recent proposal has shown promising initial results when applying SIMD instructions to parallelize \ilr{} \cite{Chen2015}. The scope of the work is however limited: (1)~it only detects faults and does not provide recovery; (2)~it only protects the floating-point unit; (3) it targets only single-threaded programs; and (4)~hardening is performed manually at the level of the program's source code. In contrast, \projectname{} targets detection \emph{and} recovery of transient CPU faults for unmodified \emph{multithreaded} programs. Furthermore, \projectname{} protects the whole CPU execution including pointers, integers, and floating-point numbers.

HAFT is a fault tolerance technique that couples \ilr{} with Hardware Transactional Memory (HTM) \cite{haft2016}. In this work, instructions are duplicated to provide fault detection, and an HTM mechanism roll-backs failed transactions to provide fault recovery. \projectname{} does not rely on a separate rollback mechanism, but rather masks faults using Triple Modular Redundancy.

Concurrent with and independent from our work, Chen et al. \cite{Chen2015,Chen2016} developed a similar approach that utilizes SIMD extensions to detect CPU faults. Solution presented in their work and \projectname{} share many similarities, though \projectname{} additionally provides recovery via triple modular redundancy and supports multithreaded applications.

\subsection{Triple Modular Redundancy}
\label{sec:nmr}

Triple Modular Redundancy (TMR) is a classical approach for achieving fault tolerance in mission-critical systems \cite{TMR1962}.
TMR detects faults by simple comparison of three replicas and performs fault recovery by majority voting, i.e., by detecting which replica differs from the other two and correcting its state.
Consequently, it imposes an obvious restriction on the fault model: only one replica is assumed to be affected by the fault.

While most of the software-based hardening techniques discussed above utilize only Dual Modular Redundancy (DMR), i.e., they can only detect but not correct faults, there are still a number of techniques based on TMR \cite{RomainMT2014,SwiftR2007}.
In the context of \ilr{}, SWIFT-R \cite{SwiftR2007} extends the fault detection mechanisms of SWIFT \cite{Swift2005} by inserting three copies (instead of two) for each instruction and performing periodic majority voting to detect and correct faults.
\projectname, in contrast, implements TMR without an increase in the number of instructions, since AVX registers are large enough to hold at least 4 copies of the data.
 
\subsection{\intelavx}
\label{sec:simd_back}

\begin{figure}[h]
\vspace{3mm}
\centering
\includegraphics[scale=0.27]{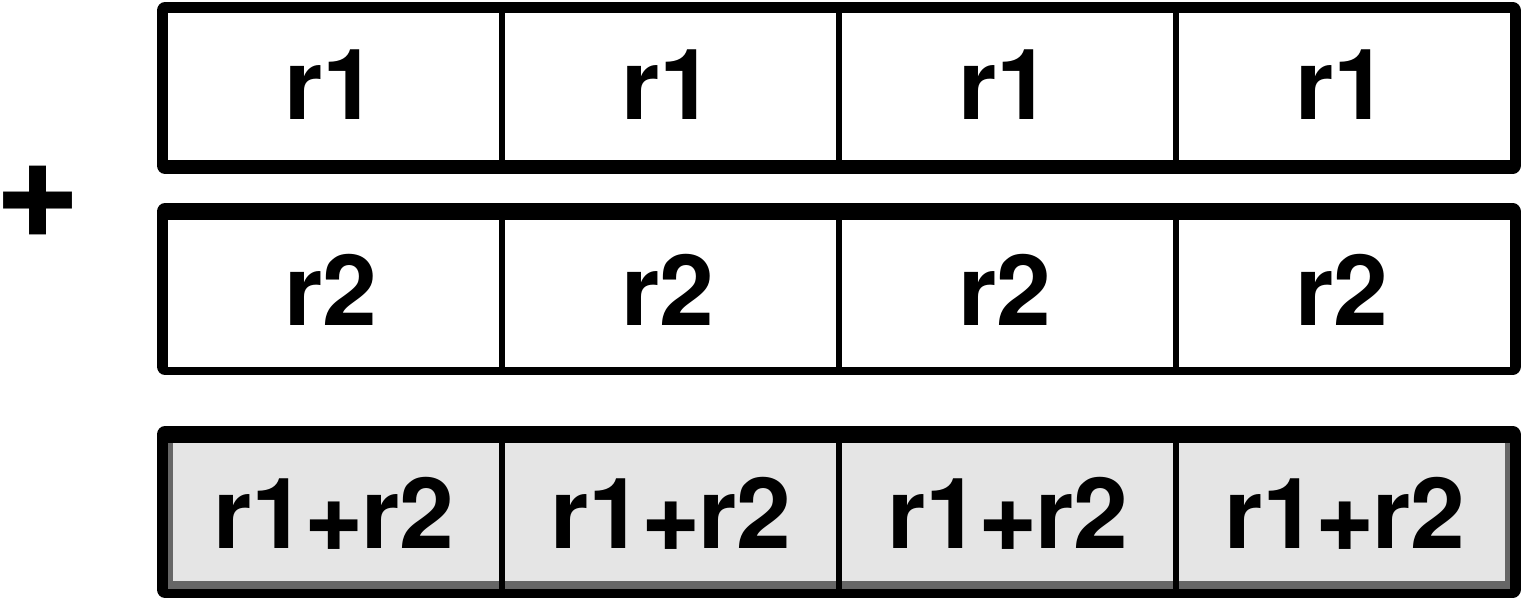}
\caption{Addition in AVX. The original values \code{r1} and \code{r2} are replicated throughout the AVX registers. All four copies are computed in parallel.}
\label{fig:avxinstruction}
\vspace{2mm}
\end{figure}

Our solution relies heavily on the Single Instruction Multiple Data (SIMD) technology and its specific implementation, \intelavx. The main idea behind it is to perform the same operation on multiple pieces of data simultaneously (data level parallelism). \figref{avxinstruction} illustrates this concept and how it relates to replication for fault tolerance. AVX adds new wider registers (YMM registers) that are capable of storing several elements and the corresponding new instructions that operate on these elements in parallel. Initially, AVX was targeted for applications that perform parallel data processing such as image or video processing; in this work, we (ab)use it for fault recovery. Note that we do not use the previous generation of Intel's SIMD implementation, SSE, since it can only operate on two 64-bit values and we need at least three copies to be able to correct faults.

\begin{figure}[t]
\vspace{2mm}
\centering
\includegraphics[scale=0.3]{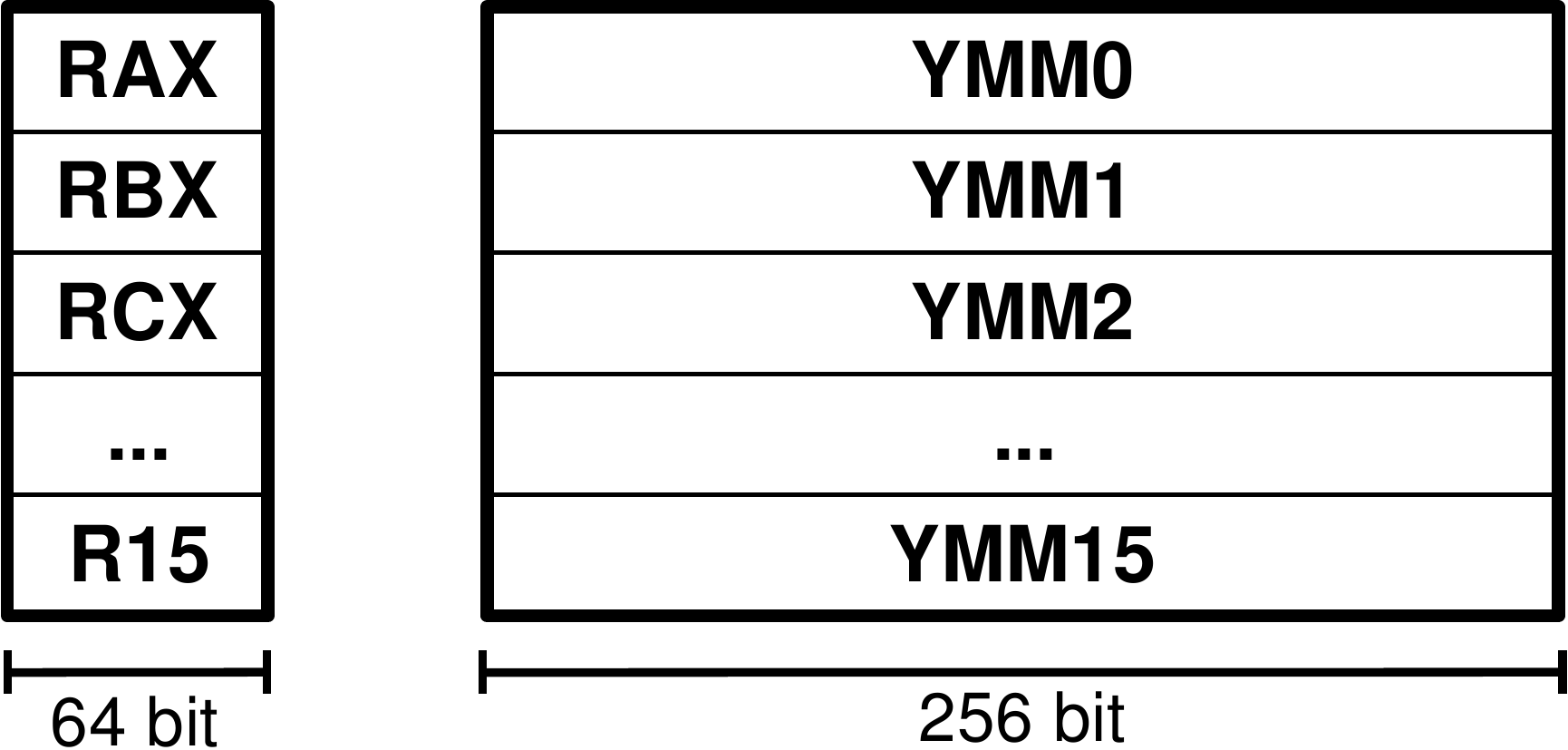}
\caption{General purpose (GPR) and AVX (YMM) registers.}
\label{fig:avx_registers}
\vspace{2mm}
\end{figure}

\myparagraph{Hardware implementation.}
The x86-64 architecture provides 16 256-bit wide YMM registers available for AVX instructions. \figref{avx_registers} compares them with general-purpose registers (GPRs). It should be noted, however, that even though only 16 registers are visible at the assembly level, many more registers are implemented physically and used at runtime (e.g., 168 YMM registers in Intel Haswell).

In modern implementations, AVX has several dedicated execution units.
It provides a high level of parallelism and allows programs to avoid some common bottlenecks.

\myparagraph{Instruction set.}
The AVX instruction set consists of a large number of instructions, including special-purpose extensions for cryptography, multimedia, etc. \projectname{} uses only a subset of AVX instructions, which we discuss in the following. 

Most arithmetic and logic operations are covered by AVX, except for integer division and modulo. For example, \figref{avxinstruction} illustrates how addition is performed with AVX.

AVX-based comparisons act differently than their counterparts in the general instruction set.
Instead of directly affecting the flags in the x86 \flagsreg{} register as normal comparisons do, AVX comparisons return either all-1 (if result is ``true'') or all-0 (``false'') values for each YMM element. This behavior is explained by the fact that the comparison is performed in parallel on multiple pieces of data, with possibly conflicting outcomes that would affect the flags differently.
On the other hand, there are no control flow instructions in the general instruction set that could operate on such sequences of 1s and 0s.
Therefore, a \code{ptest} AVX instruction was introduced that sets the ZF and CF flags in \flagsreg{} by performing an \code{and}/\code{andn} operation between its operands.\footnote{We omit the detailed explanation of how \code{ptest} works for the sake of simplicity. We refer the reader to the Intel architecture manuals.}
As a result, a branch is encoded in AVX as a sequence of an AVX comparison followed by a \code{ptest} and a subsequent jump based on the ZF and CF flags.

\begin{figure}[h]
\vspace{3mm}
\centering
\includegraphics[scale=0.3]{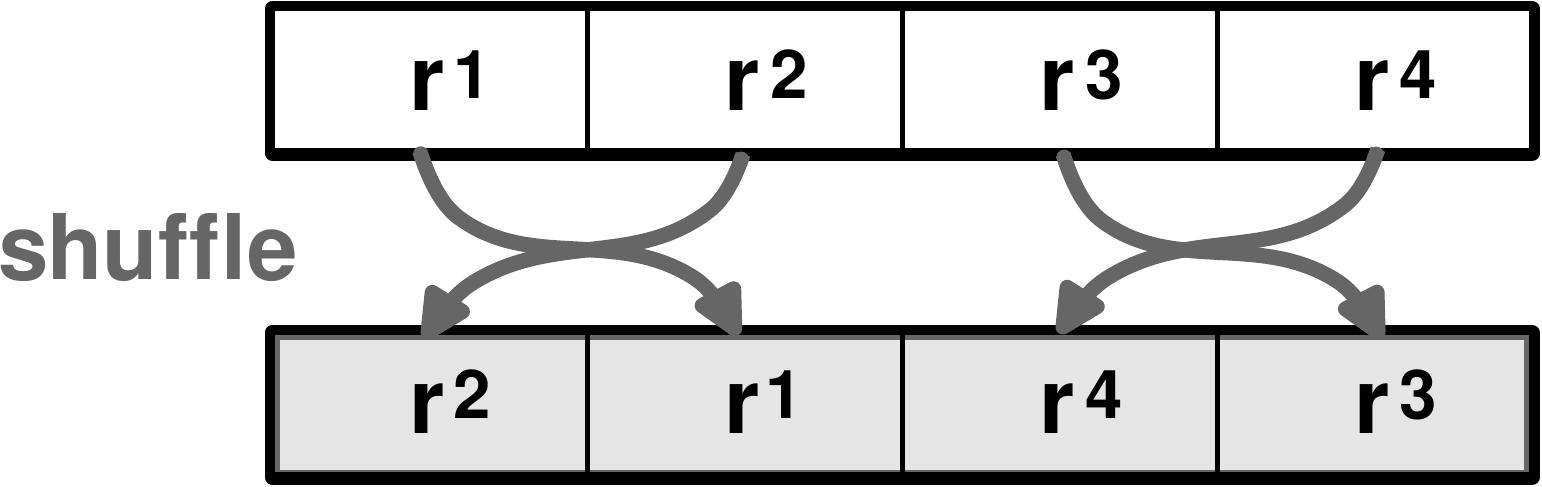}
\caption{Shuffle instruction.}
\label{fig:shuffle}
\vspace{2mm}
\end{figure} In this work, we use \code{shuffle}, a specific AVX operation that performs data rearrangement inside a YMM register. One example of a shuffle is shown in \figref{shuffle}. In combination with other operations, it allows us to get much of the functionality that is not implemented in hardware. For example, we can get a horizontal test for equality using a combination of \code{shuffle}, \code{xor} and \code{ptest} (see \secref{elzar} for more details).

 \section{Design}
\label{sec:design}

In this section, we introduce the design of \projectname{} and describe the principle of \ilr{} upon which it is based.

\subsection{System Model}
\label{sec:model}

\myparagraph{Fault model.}
\projectname{} uses the Single Event Upset (SEU) fault model~\cite{Swift2005}, where only one bit-flip in a CPU is expected to occur during the whole execution of a program. A bit-flip means an unexpected change in the state of a CPU register or a wrong result of a CPU operation. The SEU is transient, i.e., it does not permanently damage the hardware and lasts only for several clock cycles.

We fully protect the AVX register file and the AVX operations; recall that they are completely decoupled from the regular GPR registers and
scalar instructions (\secref{simd_back}).
We do not consider faults in the memory subsystem since it is assumed to be protected by ECC.
Our fault model also does not cover control flow errors, assuming some orthogonal control flow checker.

In general, \projectname{} protects from more than single faults. Indeed, four copies of data can tolerate two independent SEUs with a high probability: If any two copies agree and each of the other two copies disagree with the former ones, the majority voting can still mask the faults in the latter copies (we elaborate more on that in \secref{elzar}). In what follows, we focus on tolerating single faults for simplicity.

\myparagraph{Memory and synchronization model.}
\projectname{} imposes no restriction on the underlying memory and synchronization model,  and even works with programs containing data races. \projectname{} does not replicate nor modify the original memory-related operations (loads, stores, atomics) in any way, therefore the program's memory access behavior is unchanged. As a result, \projectname{} allows for arbitrary thread interleavings in multithreaded programs and supports all kinds of synchronization primitives.

\subsection{Instruction-Level Redundancy}
\label{sec:ilr}

\begin{figure}[t]
\vspace{-1mm}
\begin{minipage}[t]{0.26\columnwidth}
\begin{lstlisting}[name=basics1,title={(a) Native},frame=tb,framesep=0pt,aboveskip=0pt,belowskip=0pt,numbersep=2pt,numberblanklines=true,label=algo:basics1]
loop:
    r1 = add r1, r2




    cmp r1, r3


    jne loop
\end{lstlisting}
\end{minipage}\hspace{5pt}\begin{minipage}[t]{0.36\columnwidth}
\begin{lstlisting}[name=basics2,title={(b) \ilr},frame=t,framesep=0pt,aboveskip=0pt,belowskip=0pt,numbers=none,label=algo:basics2]
loop:
    r1 = add r1, r2
\end{lstlisting}
\begin{lstlisting}[name=basics2,frame=none,framesep=0pt,aboveskip=0pt,belowskip=0pt,numbers=none,backgroundcolor=\color{Lavender}]
    r1' = add r1', r2'
    r1'' = add r1'', r2''
    majority(r1, r1', r1'')
    majority(r3, r3', r3'')
\end{lstlisting}
\begin{lstlisting}[name=basics2,frame=b,framesep=0pt,aboveskip=0pt,belowskip=0pt,numbers=none]
    cmp r1, r3


    jne loop
\end{lstlisting}
\end{minipage}\hspace{5pt}\begin{minipage}[t]{0.33\columnwidth}
\begin{lstlisting}[name=basics3,title={(c) \projectname},frame=t,framesep=0pt,aboveskip=0pt,belowskip=0pt,numbers=none,label=algo:basics3]
loop:
    y1 = add y1, y2




    y4 = cmpeq y1, y3
\end{lstlisting}
\begin{lstlisting}[name=basics3,frame=none,framesep=0pt,aboveskip=0pt,belowskip=0pt,numbers=none,backgroundcolor=\color{Lavender}]
    ptest y4
    ja  recover(y4)
\end{lstlisting}
\begin{lstlisting}[name=basics3,frame=b,framesep=0pt,aboveskip=0pt,belowskip=0pt,numbers=none,rulecolor=\color{Black}]
    je loop
\end{lstlisting}
\end{minipage}
\vspace{2mm}
\caption{Original loop (a) increments r1 by r2 until it is equal to r3. Usual ILR transformation (b) triplicates instructions and adds majority voting before comparison. AVX-based \projectname{} (c) replicates data inside YMM registers, inserts \code{ptest} for comparison, and jumps to majority voting only if a discrepancy is detected in y4.}
\label{fig:basics}
\vspace{2mm}
\end{figure}
 
We base \projectname{} on Instruction-Level Redundancy (\ilr) \cite{EDDI2002,Swift2005,SwiftR2007}, a software-based technique to detect and tolerate transient hardware faults. As other software-based approaches, \ilr{} transforms the original program by replicating its computation and inserting periodic checks on computation results. An example of an \ilr-transformed code snippet is shown in \figref{basics}b.

\myparagraph{Replication.} \ilr{} replicates programs at the level of instructions. At compile-time, \ilr{} inserts ``shadow'' copies for each instruction except for a few instructions classified as ``synchronization'' instructions. The shadow copies operate on their own set of shadow registers. At runtime, the program effectively executes the original and the shadow instructions, creating mostly independent original and shadow data flows which synchronize only on specific instructions.

The synchronization instructions include all memory-related operations (loads, stores, atomics) and control-flow operations (branches, function calls, function returns). Memory-related operations are not replicated for two reasons: (a)~the memory subsystem contains only one copy of the state and there is no need to store twice, and (b)~\ilr{} keeps the memory access behavior unmodified in order to allow for non-determinism in multithreaded applications. Control-flow operations are not replicated because \ilr{} protects only data integrity and assumes no control-flow faults. Note that by not replicating function calls, \ilr{} requires no changes in function signatures and no wrappers for system calls and third-party non-hardened libraries.

To create a shadow data flow, \ilr{} replicates all inputs: values loaded from memory, values returned by function calls, and function arguments. This is achieved by a simple move of an input value in one of the shadow registers.

If only fault detection is required, it is sufficient to \emph{duplicate} the instructions and signal an error or simply crash if two data flows diverge \cite{EDDI2002,Swift2005}. If fault tolerance is needed, the instructions must be \emph{triplicated} and majority voting must be used to mask faults in one of the three data flows (see \figref{basics}b) \cite{SwiftR2007}.

\myparagraph{Checks.} To be able to detect faults, \ilr{} additionally inserts checks right before synchronization instructions. As one example, a load address must be checked before the actual load, otherwise a wrong value could be undetectably loaded and used by the subsequent instructions. As another example, all function arguments must be checked before the function call to prevent the callee from computing with wrong values. Finally, it is important to check the branch condition before branching or else the program could take a wrong path.

The checks themselves are straightforward. If crash-stop behavior is sufficient, a check compares two copies of data and crashes the program if the copies diverge. For availability (fault tolerance), \ilr{} requires majority voting on three replicas to mask a possible fault (as depicted in \figref{basics}b). During majority voting, three copies of data are compared to each other, and if one copy differs from the other two it is overwritten with the majority value.

\subsection{\projectname}
\label{sec:elzar}

As appears clearly in \figref{basics}, \ilr{} requires three times more instructions than the original program plus expensive majority voting on synchronization events. As a result, a simple 3-instruction loop may require around 13 instructions under \ilr. Such a blow-up in instructions can quickly saturate CPU resources and result in high performance overhead.

\projectname{}, on the other hand, does not replicate instructions but rather data and thus increases the total number of instructions only modestly. \figref{basics}c shows that \projectname{} inserts only 2 additional instructions to perform a check on a branch condition. The replication is achieved by utilizing wide YMM registers, with \code{y1}--\code{y4} each containing four copies of the original values. The \code{add} and \code{cmp} instructions in this snippet are actually AVX instructions which operate on four copies inside the YMM registers in parallel. The somewhat peculiar check consists of the \code{ptest} AVX instruction and a subsequent jump to recovery code if a discrepancy in branch condition \code{y4} is detected; we cover AVX-based checks in detail below.

In general, \projectname{} transforms a program as follows: it (1)~replicates the data in YMM registers, (2)~inserts periodic checks, and (3)~inserts recovery routines. In the following, we discuss each of these steps in detail.

\myparagraph{Step 1: Replication.} AVX provides an almost complete set of arithmetic and logical instructions: addition, subtraction, multiplication, bitwise operations, shifts, etc. For floating point data, all the usual instructions are present in AVX. For integers, the only missing instructions are integer division and modulo; \projectname{} falls back to basic \ilr{} in these cases. In general, \projectname{} achieves replication by simply replacing the original arithmetic and logical instructions with their AVX counterparts, as in \figref{avxinstruction}.

\begin{figure}[h]
\vspace{3mm}
\centering
\includegraphics[scale=0.35]{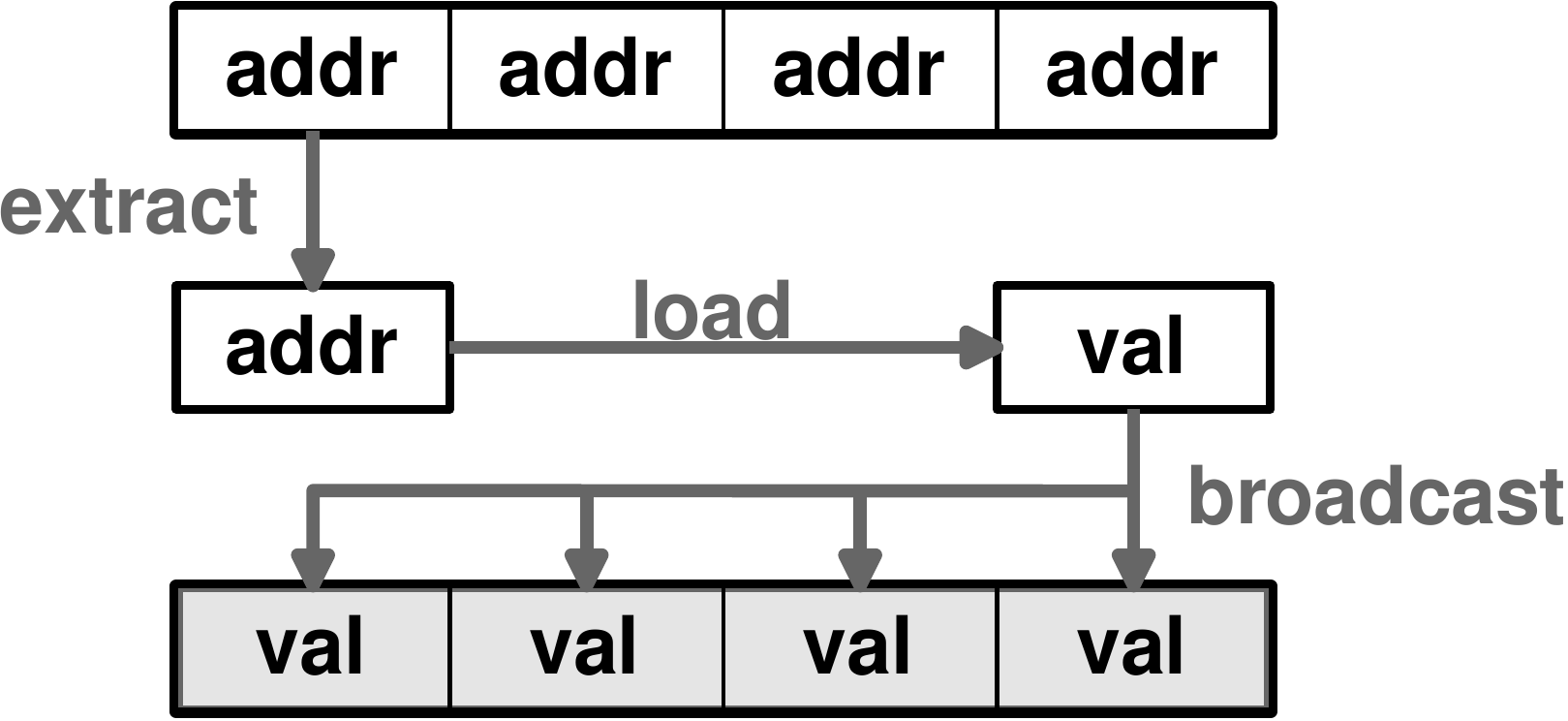}
\caption{Loads in \projectname. The original \code{load} is wrapped by AVX-based \code{extract} and \code{broadcast}.}
\label{fig:avxload}
\vspace{2mm}
\end{figure}

The situation is more complicated for (most) non-replicated synchronization instructions. These are the regular loads, stores, function calls, etc., which do not operate on YMM registers. Thus, \projectname{} has to extract one copy of each instruction's argument from YMM registers and use this copy in the instruction. If a synchronization instruction returns a value (e.g., load), this value must then be replicated inside a YMM register. AVX provides dedicated instructions for such purposes: \code{extract} and \code{broadcast}. Unfortunately, these additional instructions must wrap every single load, store, etc., which leads to high overheads. An example of such ``wrapping'' for a load is shown in \figref{avxload}.

\begin{figure}[h]
\vspace{3mm}
\centering
\includegraphics[scale=0.23]{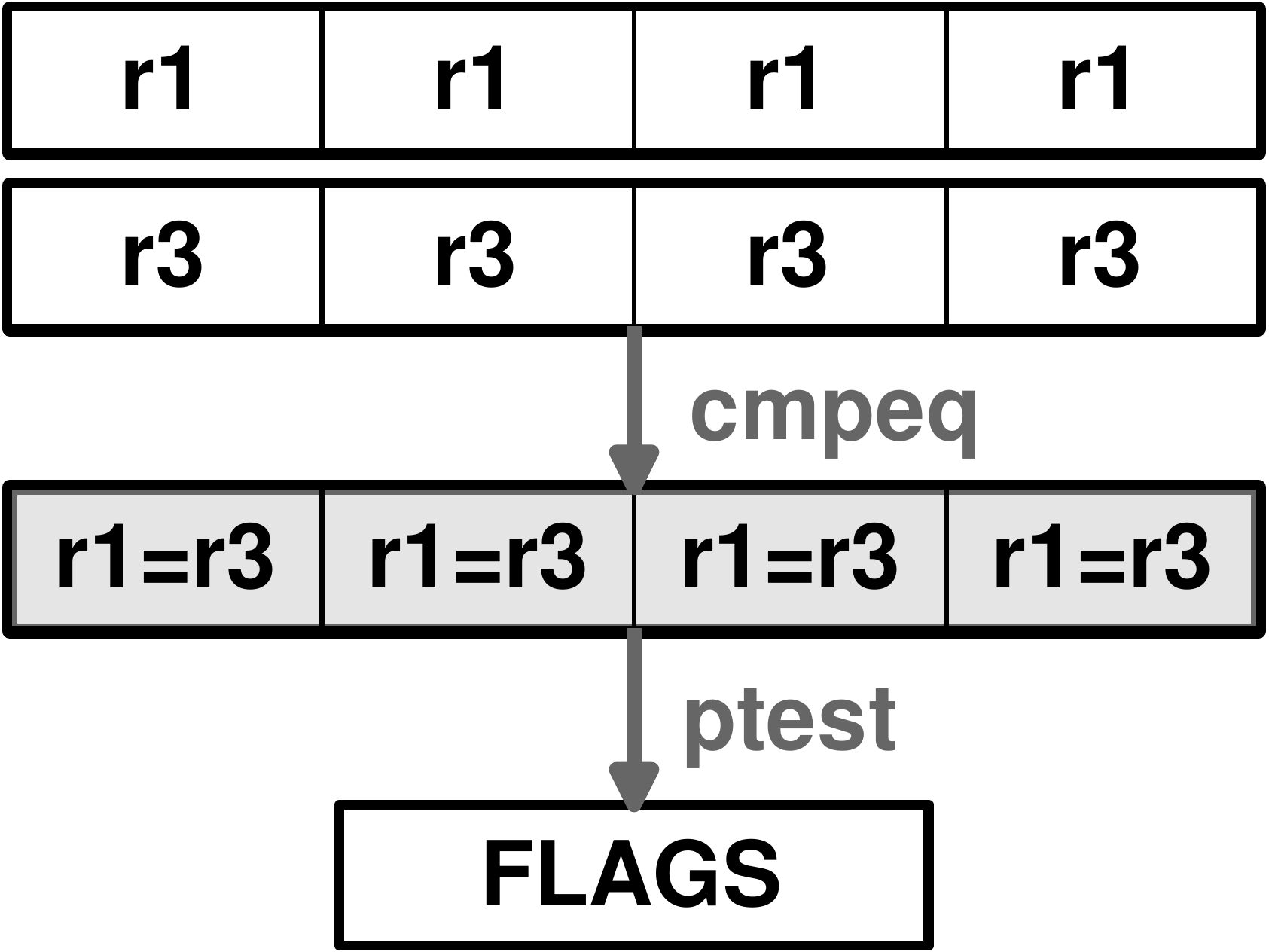}
\caption{Branching in \projectname. The original \code{cmp} for equality is transformed in a sequence of \code{cmpeq} and \code{ptest}.}
\label{fig:avxbranch}
\vspace{2mm}
\end{figure}

A special case of a synchronization instruction is a branch. A typical x86 branching sequence consists of one comparison (\code{cmp}) which toggles the \flagsreg{} register and the subsequent jump instruction (\code{je} for ``jump if equal'', \code{jne} for ``jump if not equal'', etc.). This is exemplified in Lines 7--10 of \figref{basics}a.
Unfortunately, as explained in \secref{simd_back}, AVX lacks instructions affecting control flow except for \code{ptest}. Moreover, the AVX-based comparison instructions (e.g., \code{cmpeq}) do not toggle the \flagsreg{} register but instead fill the elements of a YMM register with true/false values. Therefore, \projectname{} inserts an additional \code{ptest} to examine the result of \code{cmpeq} and only then proceeds to a jump (see \figref{avxbranch} and also \figref{basics}c, Lines 7, 8, and 10).

\myparagraph{Step 2: Adding checks.} In order to detect faults, \projectname{} inserts checks before each synchronization instruction. If a check succeeds, i.e., all copies of a YMM register contain the same value, the program continues normally, otherwise the YMM register must be recovered via majority voting. Note that the check itself must be as efficient as possible since it executes on the fast path. The recovery routine, however, resides on the slow path and can hence be less efficient.

Similar to replication, \projectname{} distinguishes between branches and all other synchronization instructions. Because of implementation choices in AVX, checks turn out to be very effective for branches but not for other operations. To support efficient checks in \projectname{}, we rely on the assumption that a fault corrupts only one copy in a YMM register (see \secref{model}).

\begin{figure}[h]
\vspace{3mm}
\centering
\includegraphics[scale=0.35]{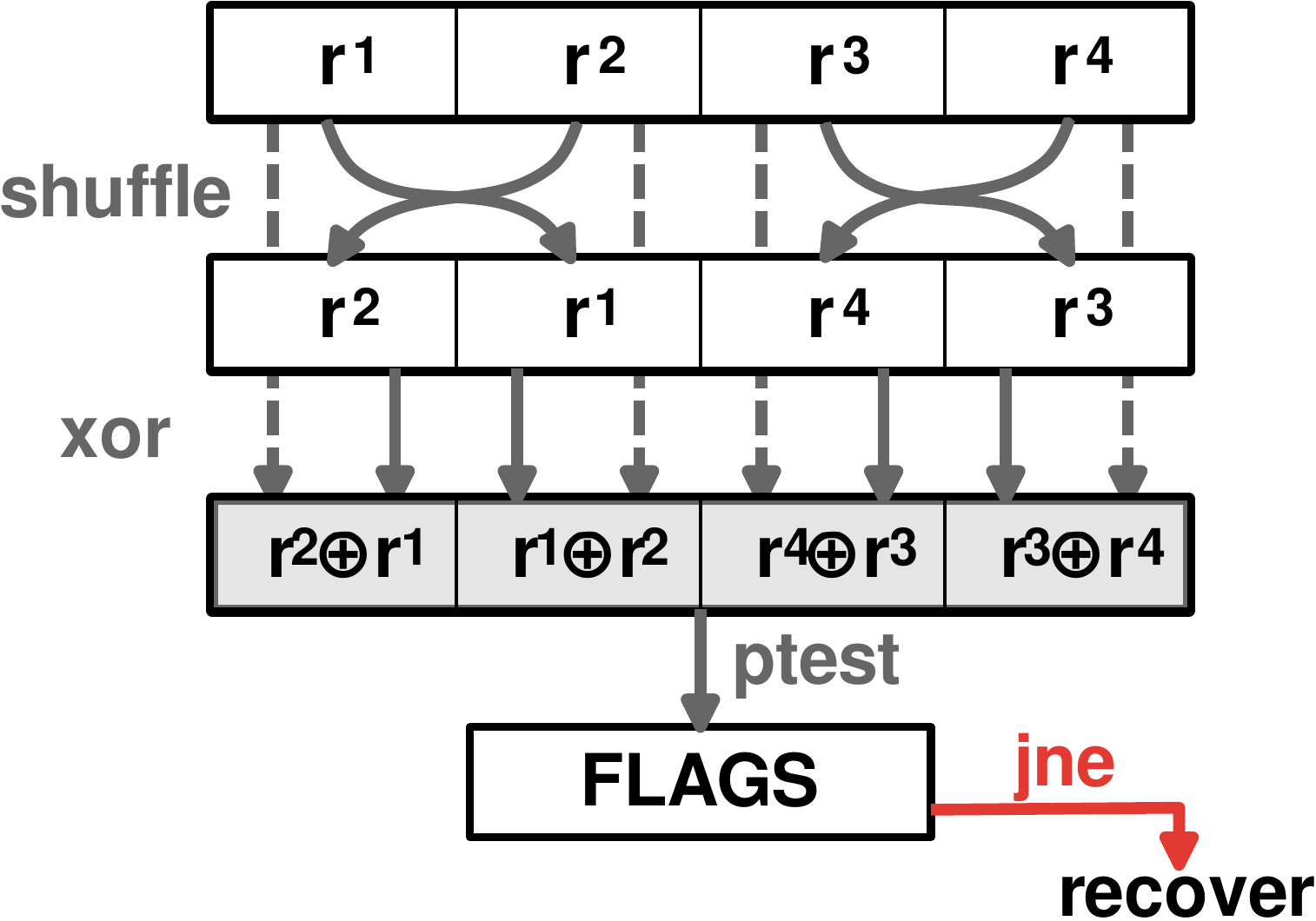}
\caption{Check on synchronization instructions except branches (loads, stores, etc.). \projectname{} \code{shuffle}s a YMM register, \code{xor}s the shuffled register with the original, checks the result (\code{ptest}), and jumps to a recovery routine (\code{jne}) in case of error.}
\label{fig:avxcommoncheck}
\vspace{2mm}
\end{figure}

In general, a check on the arguments of a synchronization instruction requires a pair-wise (horizontal) comparison of copies inside a YMM register. For example, upon a function call, all function arguments replicated in the corresponding AVX registers must be checked for discrepancies. Interestingly, AVX provides a horizontal subtraction instruction called \code{hsub}, but it is not implemented for 64-bit integers and is generally slow. Hence, we opted for another implementation of checks that involves a \code{shuffle} and a subsequent \code{xor}. This idea is illustrated in \figref{avxcommoncheck}. In an error-free case, \code{xor} produces all-$0$s which is easily ruled out by \code{ptest}. In the case of a fault in one of the copies, the result of \code{xor} is a mix of $0$s and $1$s, which triggers the \code{jne} path and leads to recovery.

\begin{figure}[h]
\vspace{3mm}
\centering
\includegraphics[scale=0.35]{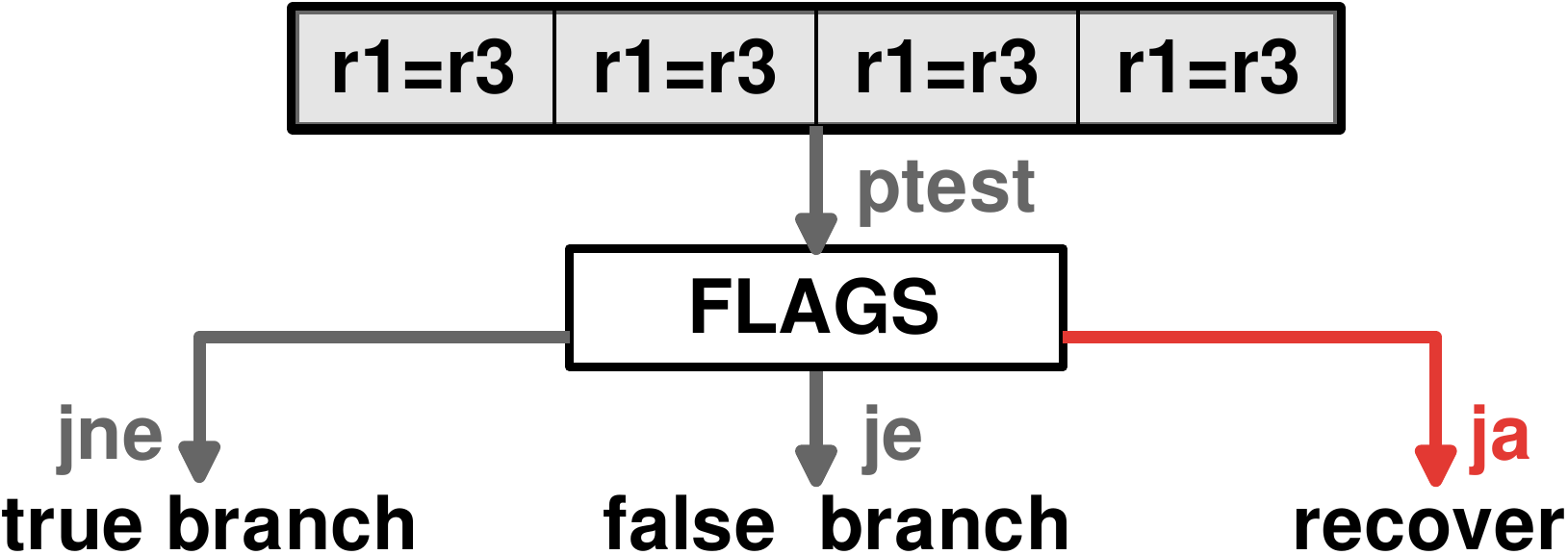}
\caption{AVX-based \code{ptest} toggles \flagsreg{} such that all-true and all-false outcomes respectively correspond to the true and false branches, whereas a mix of true and false indicates a fault and triggers recovery.}
\label{fig:avxbranchcheck}
\vspace{2mm}
\end{figure}

A check on a branch is cheaper and conceptually simpler. As evident in \figref{avxbranch}, branching in AVX already requires an AVX-based comparison and a \code{ptest}. We notice that in error-free case, comparisons in \projectname{} can produce only all-true or all-false results (see \secref{simd_back}). Thus, a mix of true and false indicates a fault. Fortunately, \code{ptest} is a versatile instruction that allows us to check for an all-true, all-false, or true-false mix outcome simultaneously, as shown in \figref{avxbranchcheck}. Therefore, to add a check before a branch, it is sufficient to augment the AVX-based branching with just a single jump instruction, \code{ja} (``jump if above''), as shown in \figref{basics}c, Line 9.

\myparagraph{Step 3: Adding recovery.} Checks on branches and other synchronization instructions trigger a recovery routine when a fault is detected. The task of this routine is to mask a fault.
Because of the assumption that a fault is localized in only one copy of the YMM register (see \secref{model}), it is sufficient to identify two identical replicas in the register and blindly broadcast their value to the whole register.
This can be performed efficiently by a single comparison of the low elements of the faulty YMM register (depicted in gray in \figreftwo{avxcommoncheck}{avxbranchcheck}) and, depending on the result of the comparison, copying either the lowest or the highest element to the rest of the register.

We note, however, that we can easily implement a smarter recovery strategy that would support more complex fault patterns involving multiple bit flips.
As the recovery procedure is on the slow path, i.e., it is triggered only rarely, it does not need to be optimized for speed and this added reliability can be implemented without compromising performance.

The idea of the extended recovery procedure is to check all four elements and consider three scenarios:
(1)~if three elements are identical, then the last one is faulty and can be overwritten with the value of the former;
(2)~if two elements are identical and the other two have each a different value, then the latter elements are both faulty and can be overwritten with the value of the former;
finally, (3)~if we have two groups of two elements, with each group agreeing on a different value, then the same fault has affected two elements and we have no majority, hence program execution must stop.
This recovery strategy can tolerate all single bit flips, all flips of two bits of different order in the replicas, as well as a wide variety of more complex fault patterns that leave at least two elements identical.

\subsection{Data Types Support}
AVX natively supports 8-, 16-, 32-, and 64-bit integers as well as single- and double-precision floating points. However, up to this moment the discussion implied 64-bit integers replicated four times across a 256-bit YMM register.

There are three options to support smaller types: (1)~cast all smaller integer types to 64-bit integers and 32-bit floats to 64-bit doubles, (2)~replicate all types only four times in the low bits of YMM registers, leaving upper bits nullified, or (3)~replicate smaller types so many times as to fill up the whole YMM register. The first approach obviously breaks semantics of integer overflows and floating point precision, possibly leading to unexpected computation results. The second approach is better but requires additional care for AVX instructions that compute across the whole YMM register, e.g., results of comparisons may differ in lower and upper bits. Therefore we chose the third approach which leads to extreme settings of up to 32-modular redundancy for 8-bit integers but is conceptually clean.

Compilers like LLVM sometimes produce esoteric integer types like 1-bit or 9-bit integers, usually for sign-extension and truncation purposes. Such data types are rare but still present in many applications, therefore we extend them to the AVX-supported bit width and treat them as ``usual'' integers. We take special care whether to zero- or sign-extend them, depending on the associated semantics.

 \section{Implementation}
\label{sec:impl}

We implemented \projectname{} as an LLVM compiler pass \cite{LLVM2004} that takes unmodified source code of an application and emits an AVX-hardened executable. We also implemented a fault injection framework to be able to test \projectname's fault tolerance capabilities. 
\subsection{Compiler Framework}
\label{sec:compiler_impl}

\myparagraph{Tool chain.} We developed \projectname{} as a compiler pass in LLVM 3.7.0 ($\sim600$ LOC). Additionally, we extract the implementation of checks and recovery in a separate LLVM IR file ($\sim250$ LOC). This separation allowed us to write the pass in a (mostly) target-independent way, i.e., AVX can be substituted by another similar technology (e.g., ARM Neon) by rewriting only the IR file with checks and recovery.

\projectname{} is plugged in the usual build process of an application, i.e., there is no need to modify the source code or the makefiles/configuration scripts. To achieve this, we employ the LLVM gold linker plugin that can save the final optimized and linked LLVM bitcode to a file. \projectname{} takes this file as input, adds AVX-based redundancy, and emits the hardened executable. Thus, \projectname{} performs its transformation after all optimization passes and right before assembly code generation.

In order to be able to use AVX for replication, we disallow any vectorization in original programs.
All other optimizations are enabled. Additionally, we run the \emph{scalarrepl} pass to replace all aggregate data types (structs, arrays) because they are not natively supported by LLVM vectors we employ.

\myparagraph{Pass details.} The usual way to write AVX-enabled programs is to use AVX intrinsics or directly AVX inline assembly. This approach is the closest to ``bare metal'' and allows for fine performance tuning, but it is also time-consuming and error-prone. Moreover, using intrinsics or inline assembly would make it impossible to directly port \projectname{} to a different technology than Intel AVX.

Fortunately, LLVM provides first-class \emph{vector} types that were specifically introduced for SIMD programming and come with an extensive support for vector operations. The x86 code generator recognizes vectors and transforms them into AVX instructions. LLVM also introduces three special instructions to work with vectors, \code{extractelement}, \code{insertelement}, and \code{shufflevector} that are respectively mapped to AVX's \code{extract}, \code{broadcast}, and \code{shuffle}. Generally, we found vectors to be a very powerful abstraction, with the quality of the generated AVX code improving with each LLVM release.

With LLVM vectors, the process of AVX hardening becomes fairly trivial: (1)~all data types of a program are transformed into corresponding vector types, (2)~each of the synchronization instruction's arguments is extracted from a vector using \code{extractelement}, (3)~each synchronization instruction's return value is broadcast to the whole vector using \code{insertelement}, (4)~all other instructions are substituted to work on the corresponding vectors, and (5)~checks and recovery routines are inserted before synchronization instructions. An example of \projectname-transformed program is shown in \figref{llvm}.

\begin{figure}[t]
\centering
\begin{minipage}[t]{0.39\columnwidth}
\begin{lstlisting}[style=llvm, name=llvm1,title={(a) Native},frame=tb,framesep=0pt,aboveskip=0pt,belowskip=0pt,numbersep=2pt,numberblanklines=true,label=algo:basics1]
loop:
    r1 = add i64 r1, r2
    c = cmp eq i64 r1, r3
    
    
    
    br i1 c, exit, loop
\end{lstlisting}
\end{minipage}\hspace{5pt}\begin{minipage}[t]{0.51\columnwidth}
\begin{lstlisting}[style=llvm, name=llvm2,title={(b) \projectname},frame=t,framesep=0pt,aboveskip=0pt,belowskip=0pt,numbers=none,label=algo:basics2]
loop:
    r1 = add |\color{darkgray}\textbf{<4 x i64>}| r1, r2
    c1 = cmp eq |\color{darkgray}\textbf{<4 x i64>}| r1, r3
\end{lstlisting}
\begin{lstlisting}[style=llvm, name=llvm2,frame=none,framesep=0pt,aboveskip=0pt,belowskip=0pt,numbers=none,backgroundcolor=\color{Lavender}]
    c64 = sext c1 to |\color{darkgray}\textbf{<4 x i64>}|
    t = call ptest(|\color{darkgray}\textbf{<4 x i64>}| c64)
    c = cmp eq i32 t, 0
\end{lstlisting}
\begin{lstlisting}[style=llvm, name=llvm2,frame=b,framesep=0pt,aboveskip=0pt,belowskip=0pt,numbers=none]
    br i1 c, exit, loop
\end{lstlisting}
\end{minipage}\vspace{-1mm}
\caption{Example from \figref{basics} as represented in simplified LLVM IR. Original code (a) operates on \code{i64} 64-bit integers. \projectname{} (b) transforms the code to use \code{<4 x i64>} vectors of four integers. Since LLVM-based comparisons do not directly map to AVX, \projectname{} inserts some boilerplate code (shown in gray).}
\label{fig:llvm}
\vspace{3mm}
\end{figure}
 
A nice feature of this vector-based approach is that one can abstract away from the underlying AVX implementation. As such, we do not need to care about most corner cases like vector-based integer division which is not implemented in AVX. We can still write it in an LLVM vector form, and the x86 code generator automatically converts it to four regular division instructions.

The careless use of vectors, however, may seriously hamper performance in some cases. For example, a straightforward implementation of branches with LLVM vectors results in a convoluted and ineffective instruction sequence; this is related to the fact that \projectname{} uses \code{ptest} in an unusual manner that was not anticipated by the developers of the x86 code generator and is not efficiently supported in the pattern-matching rules. For such corner cases, we explicitly insert boilerplate code patterns as shown in gray in \figref{llvm}b. This code actually generates the \code{ptest}-\code{je} instruction sequence in the final executable, exactly as in \figref{basics}c.\footnote{To construct the boilerplate LLVM code, we consulted the source code of LLVM codegen's regression tests. These tests gave us a good understanding of how specific LLVM constructs are mapped to AVX assembly. This was literally a ``test-driven development'' experience.}

As discussed previously (\secref{elzar}), AVX natively supports only 8-, 16-, 32-, and 64-bit integers and 32- and 64-bit floating points. Since LLVM sometimes produces types with unsupported widths, we have no other choice but to extend them to supported types. In the case of integers, we take special care to sign- or zero-extend them. In some other cases (e.g., for SQLite3), we had to switch off the long-double type using predefined macros in the source code.

\myparagraph{Libraries support.} Most previous research in the area of \ilr{} focused on hardening only the program's source code and left third-party libraries unprotected \cite{Swift2005,SwiftR2007,Shoestring2010,ESoftCheck2009}. This leads to better performance but also to lower fault coverage, because a fault in library code can go undetected. We notice however that many programs from the Phoenix and PARSEC benchmark suites, which are used in our evaluation, heavily utilize the standard C (libc) and math (libm) libraries. Therefore, to report more accurate numbers, we also harden a significant part of libc and libm. We decided not to harden the I/O, OS, and pthreads-related functions for our prototype implementation because their execution takes less than $\sim5\%$ of the overall time. As a reference implementation, we chose the musl library with inline assembly disabled.

\myparagraph{Limitations.} Our prototype does not support inline assembly because LLVM treats assembly code as calls to undefined functions and provides no information about such code. Furthermore, our prototype does not have support for C++ exceptions.

\subsection{Fault Injection Framework}
\label{sec:fi_impl}

For time budget reasons, we ran our fault injection experiments on a medium-sized cluster of computers without AVX installed.
We therefore needed a fault injection tool that can emulate Intel AVX. Since available tools do not provide such support, we developed our own binary-level fault injector ($\sim320$ LOC) using Intel Software Development Emulator (SDE), which provides support for AVX instructions and gdb debugger.
In the following, we give a high-level overview of our fault injector.

Basically, a fault injection campaign for each program proceeds in two steps.
First, a program instruction trace is collected via the Intel SDE debugtrace tool. This preparatory step is required to automatically find and demarcate the boundaries of the hardened part of the program (remember that \projectname{} does not harden external libraries and we do not want to inject faults into them). Knowing these boundaries, our fault injection tool can narrow down the set of instructions in which the fault can be injected.

Second, the program is executed repeatedly and, in each run, a single fault is injected (\secref{model}). To that end, a program-under-test is started under Intel SDE with a gdb process attached. To inject a fault, we dynamically create a new gdb script that sets a random breakpoint for a given occurrence of a particular instruction (otherwise gdb would always stop at the first occurrence of the instruction). When the program runs under Intel SDE with gdb attached, it stops at the breakpoint, the fault injection happens, and the now-faulty program continues execution. After the program terminates, our fault injection tool examines the program output, assigns a corresponding outcome (see below), and proceeds to another fault injection run.

Each fault injection run results in one of the outcomes listed in \tabref{fi_results}. To distinguish between the correct and corrupted system states, each program-under-test is run first without fault injections to produce a reference output (``golden run''). Consequently, after each run, the program output is compared against this reference output, and a SDC is signaled if two outputs differ.

\begin{table}[t]
\renewcommand{\arraystretch}{0.8}
\footnotesize
\centering

\begin{tabular}{l p{4.0cm} p{1.2cm}}
\bfseries FI outcome & \bfseries Description & \bfseries System \\
\hline                    
\hline
\noalign{\vskip 1pt}
Hang           & Program became unresponsive & \multirow{2}{*}{Crashed} \\
OS-detected         & OS terminated program &  \\ 
\hline\\ [-6pt] 
\projectname-corrected      & \projectname{} detected and corrected fault & \multirow{2}{*}{Correct} \\
Masked         & Fault did not affect output & \\
\hline\\ [-6pt]   
SDC            & Silent data corruption in output & Corrupted \\
\hline
\end{tabular}

\caption{Fault injection outcomes classified.}
\label{tab:fi_results}
\vspace{3mm}
\end{table}
 
We inject faults by overwriting an output register of an instruction where the breakpoint was set.
We inject not only in AVX (YMM) registers but also in regular (GPR) registers.
For YMM registers, we inject faults only in one element of the register to match our fault model (\secref{model}).
 \section{Evaluation}
\label{sec:eval}

\begin{figure*}[t]
\centering
\includegraphics[scale=0.7]{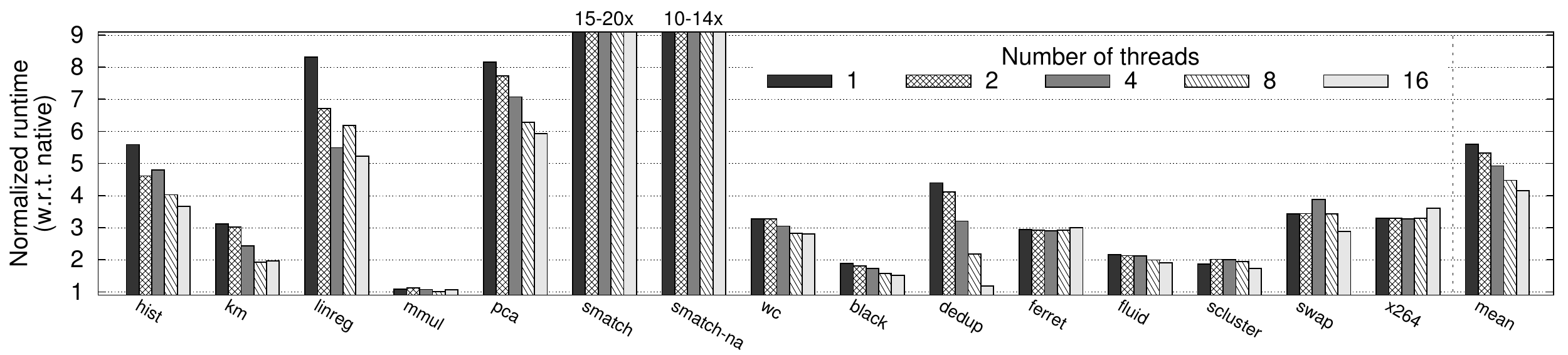}
\vspace{-2mm}
\caption{Performance overhead over native execution with the increasing number of threads.
}\label{fig:benches-scalability}
\vspace{0mm}
\end{figure*}
 
\begin{figure*}[t]
\centering
\includegraphics[scale=0.7]{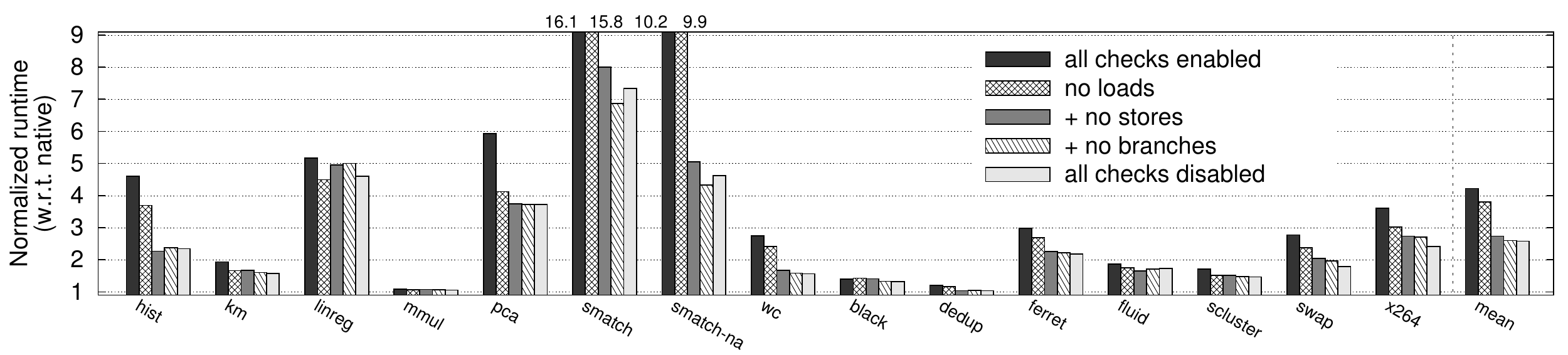}
\vspace{-2mm}
\caption{Performance overheads breakdown by disabling checks (with 16 threads).
}\label{fig:benches-checks-impact}
\vspace{0mm}
\end{figure*}
 
In this section, we answer the following questions:
\begin{itemize}
\item What is the performance overhead incurred by \projectname, and what are the causes for high overheads (\secref{perf_eval})?
\item How many faults are detected and corrected by \projectname{} during fault injection experiments (\secref{fi_eval})?
\item How does \projectname{} perform compared to a state-of-the-art \ilr{} implementation (\secref{comp_eval})?
\end{itemize}

\subsection{Experimental Setup}
\label{sec:setup_eval}

\myparagraph{Applications.} \projectname{} was evaluated on two benchmark suites: Phoenix 2.0 \cite{Phoenix2007} and PARSEC 3.0 \cite{Parsec2009}. Results are reported for all 7 Phoenix benchmarks
and 7 out of 13 PARSEC benchmarks.
The remaining 6 benchmarks from the PARSEC suite were not evaluated for the following reasons: \emph{bodytrack} and \emph{raytrace} use C++ exceptions not supported by \projectname, \emph{facesim} crashes with a runtime error when built with LLVM, \emph{freqmine} is based on OpenMP and does not compile under our version of LLVM, \emph{canneal} has inline assembly and \emph{vips} has long-double floats not supported by \projectname.

All applications were built with LLVM 3.7.0 and \projectname{} as described in \secref{compiler_impl}. The native versions were built with \code{msse4.2} and \code{mavx2} flags to enable SIMD vectorization. The \projectname{} versions were built with all vectorization disabled, i.e., with \code{no-sse}, \code{no-avx}, \code{fno-vectorize}, and \code{fno-slp-vectorize} flags. For all versions, all other compiler optimizations were enabled (\code{O3} flag). Additionally, we used the \code{fno-builtin} flag to transparently link against our versions of libc and libm.

Note that we compare \projectname{} against the native version with all AVX optimizations enabled. As \figref{benches-vectorization-impact} indicates, most benchmarks do not benefit from AVX. However, \emph{string match} shows a $60\%$ increase in performance. Therefore, we decided to also show how \projectname{} performs in comparison to the native version with AVX optimizations disabled; we refer to this experiment as \emph{smatch-na} (for ``string match no AVX'').

\myparagraph{Datasets.} For the performance evaluation, we use the largest available datasets provided by Phoenix and PARSEC. However, for the fault injection experiments, we use the smallest available inputs due to the extremely slow fault injection runs.

\myparagraph{Testbed}. The performance evaluation was done on a machine with two 14-cores Intel Xeon processors operating at 2.0~GHz (Intel Haswell microarchitecture\footnote{We also performed experiments on Intel Skylake but the results were similar to Intel Haswell. Therefore, we omit them in our evaluation.}) with 128~GB of RAM, a 3.5~TB SATA-based SDD, and running Linux kernel 3.16.0. Each core has private 32~KB L1 and 256~KB L2 caches, and 14 cores share a 35~MB L3 cache. For performance measurements, we report an average of 10 runs.

For fault injections, we used a cluster of 25 machines to parallelize the experiments. We injected a total of $2,500$ faults in each program.
All programs-under-test were run with two threads to account for the impact of multithreading.

\subsection{Performance Evaluation}
\label{sec:perf_eval}

\myparagraph{Impact of \projectname{} and scalability.} The performance overheads incurred by \projectname{} are shown in \figref{benches-scalability}. There is significant variability in behavior across benchmarks, with some showing overheads as low as $10\%$ (\emph{matrix multiplication}) and some exhibiting up to $20\times$ worse performance (\emph{string match}). On average, the normalized runtime of \projectname{} is $4.1$--$5.6\times$ depending on the number of threads.

For some benchmarks, there is also variability across the number of threads. Ideally, if a program has linear scalability, \projectname{} should incur exactly the same performance overhead with any number of threads, e.g., as in case of \emph{word count} or \emph{ferret}. However, some benchmarks such as \emph{dedup} are well-known to have poor scalability, i.e., with many threads they spend a lot of time on synchronization \cite{ParsecPerf2009}. Thus, \projectname's overhead is partially amortized by the sub-linear scalability of these benchmarks.

To gain better understanding on the causes of high overheads as well as the causes of high variability across benchmarks, we gathered runtime statistics for native and \projectname{} versions of all benchmarks. The results are shown in \tabreftwo{benches-stats}{avxbenches-stats}. The benchmarks were run with 16 threads (and in the case of \projectname, with all checks enabled) and profiled using perf-stat to collect hardware counters of raw events such as the number of loads, stores, branches, all instructions and AVX instructions only, etc.

Based on the information from \tabreftwo{benches-stats}{avxbenches-stats}, we can highlight several causes of high performance overheads. Firstly, as \tabref{avxbenches-stats} shows, \projectname{} leads to an increase in the total number of executed instructions of $4$--$8\times$ on average. This disappointingly high number is explained by the fact that \projectname{} adds wrapper instructions for loads, stores, and branches, as well as expensive checks on synchronization instructions (see \secref{elzar}).

Second, looking at the achieved Instruction-Level Parallelism (ILP) in \tabref{avxbenches-stats}, we notice that current x86 CPUs provide much better parallelization for regular instructions as compared to AVX instructions. As one example, \emph{linear regression} achieves a high ILP of $6.51$ instructions/cycle in native execution, but the AVX-based version reaches only a disappointing ILP of $1.7$. Combined with the $10.49\times$ increase in number of instructions for the AVX-based version, it is no surprise that \emph{linear regression} exhibits an overhead of $\sim5$--$8\times$.

Two benchmarks that show the lowest overheads are \emph{matrix multiplication} and \emph{blackscholes}. In the case of \emph{matrix multiplication}, almost all of \projectname's overhead is amortized by a very poor memory access pattern that leads to $62.39\%$ of all memory references missing L1 cache; in other words, \emph{matrix multiplication} spends more time in waiting for memory than in actual computation.
In the case of \emph{blackscholes}, the main cause for low overheads is the small fraction of loads/stores ($12.22\%$) and branches ($15.63\%$).

Finally, we inspected the causes for extremely high overheads in \emph{string match}. First of all, \emph{string match} by itself significantly benefits from AVX vectorization (see \figref{benches-vectorization-impact}). Indeed, \projectname{} is $\sim15$--$20\times$ slower than the native version, but $\sim10$--$14\times$ slower than native with AVX vectorization disabled. Second of all, \projectname{} increases the total number of executed instructions by a factor of $32$. Upon examining the source code of \emph{string match}, we noticed that it spends most of the time in \code{bzero} to nullify some chunks of memory. LLVM produces a very effective assembly for this helper routine, but \projectname{} inserts wrappers and checks for the store and branch instructions in \code{bzero}, leading to much longer and slower assembly code.

\begin{table}[t]
\renewcommand{\arraystretch}{0.8}
\footnotesize
\centering
\begin{filecontents*}{benches-stats.csv}
bench,ilp,lonedmissratio,branchmissratio,loadsratio,storesratio,branchesratio
hist,1.59, 0.66, 0.01,53.21,26.67, 9.56
km,3.48, 1.48, 0.33,20.83, 0.48,14.96
linreg,6.51, 2.05, 0.01,18.02, 0.21, 3.82
mmul,0.22,62.39, 0.14,40.16, 0.07,10.10
pca,2.61,12.19, 0.27,14.21, 0.21, 3.79
smatch,2.38, 0.12, 0.70,11.61,14.35,22.40
wc,1.31,10.94, 3.31,29.75,23.63,13.67
black,1.83, 0.40, 1.21, 9.38, 2.84,15.63
dedup,1.04, 4.30, 3.80,30.08,13.55,12.01
ferret,1.11, 4.69,12.65,14.47, 2.28,17.42
fluid,1.22, 1.17,14.70,11.77, 2.58,14.29
scluster,0.68, 4.17, 1.47,32.60, 0.43, 9.33
swap,1.97, 0.82, 0.97,30.98, 4.80,11.05
x264,2.11, 0.34, 0.31,26.83, 8.32,21.00
\end{filecontents*}
\csvreader[tabular=l|rr|rrr,
    table head=\bfseries Bench & \bfseries L1-miss & \bfseries br-miss & \bfseries loads & \bfseries stores & \bfseries branches \\\hline\hline\noalign{\vskip 1pt},
    table foot=\hline]
{benches-stats.csv}
{bench=\bench,ilp=\ilp,lonedmissratio=\lonedmissratio,branchmissratio=\branchmissratio,loadsratio=\loadsratio,storesratio=\storesratio,branchesratio=\branchesratio}
{\bench & \lonedmissratio & \branchmissratio & \loadsratio & \storesratio & \branchesratio}
\caption{Runtime statistics for native versions of benchmarks with 16 threads: L1D-cache and branch miss ratios, and fraction of loads, stores, and branches over executed instructions (all numbers in percent).}\label{tab:benches-stats}
\vspace{2mm}
\end{table}

\begin{table}[t]
\renewcommand{\arraystretch}{0.8}
\footnotesize
\centering

\begin{tabular}{l | r r r | r r}
 & \multicolumn{3}{c|}{\bfseries Instruction-Level Parallelism} & \multicolumn{2}{c}{\bfseries Increase in \# of instr} \\
 & \multicolumn{3}{c|}{\bfseries (ILP), instr/cycle} & \multicolumn{2}{c}{\bfseries w.r.t. native} \\
\bfseries Bench & \bfseries Native & \bfseries \projectname & \bfseries \swiftr & \bfseries \projectname & \bfseries \swiftr \\
\hline
\hline
\noalign{\vskip 1pt}
hist      & 1.59  & 2.13  & 4.30 &  8.56 &  6.17 \\
km        & 3.48  & 2.58  & 3.85 &  6.37 &  4.34 \\
linreg    & 6.51  & 1.70  & 3.46 & 10.49 &  4.33 \\
mmul      & 0.22  & 0.96  & 1.71 &  4.47 &  7.77 \\
pca       & 2.61  & 2.28  & 3.89 &  6.82 &  9.45 \\
smatch    & 2.38  & 3.26  & 3.46 & 32.72 & 11.56 \\
wc        & 1.31  & 2.24  & 3.05 &  6.14 &  3.42 \\
black     & 1.83  & 1.77  & 2.97 &  1.70 &  5.18 \\
dedup     & 1.04  & 1.75  & 2.00 &  4.64 &  3.68 \\
ferret    & 1.11  & 1.81  & 2.57 &  4.32 &  6.33 \\
fluid     & 1.22  & 1.54  & 2.77 &  2.43 &  6.02 \\
scluster  & 0.68  & 1.22  & 1.34 &  3.77 &  3.87 \\
swap      & 1.97  & 2.06  & 2.68 &  3.50 &  4.40 \\
x264      & 2.11  & 2.00  & 3.44 &  3.26 &  3.71 \\
\hline
\end{tabular}

\caption{Runtime statistics for \projectname{} and \swiftr{} versions of benchmarks with 16 threads: Instruction-Level Parallelism (ILP) and increase factor in the number of executed instructions w.r.t. native.}\label{tab:avxbenches-stats}
\vspace{2mm}
\end{table}
 
\myparagraph{Impact of checks.} We also investigated the impact of checks inserted by \projectname{} (see \secref{elzar}). \figref{benches-checks-impact} shows the results of successively disabling checks on loads, stores, branches, and all other instructions (e.g., function calls, function returns, atomics). Note that the results are shown for benchmarks run with $16$ threads.

We observe that checks constitute a significant part of the overall performance overhead of \projectname. For example, disabling checks on loads and stores decreases the overhead from $4.2$ to $2.7\times$ on average, a difference of $55\%$. Disabling checks on branches leads to a negligible overhead reduction of $4\%$, which proves that our branch checking scheme is very efficient (\secref{elzar}).

We also observe that disabling checks on loads and stores respectively reduces the overhead by $11\%$ and $40\%$, i.e., checks on stores have higher overheads than checks on loads. The reason is that stores require to check both the address and the value to store whereas loads only need to check the address.

\myparagraph{Floating point-only protection.} As AVX was initially developed to accelerate floating-point calculations, it is interesting to study the overheads when applying \projectname{} only to floating-point data. We thus developed a stripped-down version of \projectname{} that replicates floats and doubles but not integers and pointers, and ran tests on several PARSEC benchmarks that contain sufficiently many floating-point operations: \emph{blackscholes} ($47\%$ of all instructions are floating-point), \emph{fluidanimate} ($32\%$), and \emph{swaptions} ($34\%$) \cite{Parsec2009}.

Our results prove that \projectname{} hardens floating points with a low overhead. Depending on the number of threads, we observe a $9$--$35\%$ performance overhead over native for \emph{blackscholes},\footnote{This is in line with the numbers reported by Chen et al. \cite{Chen2015} where a single-threaded, manually written SSE-based version of \emph{blackscholes} exhibits $\sim30\%$ overhead.} $10$--$18\%$ for \emph{fluidanimate}, and $40$--$60\%$ for \emph{swaptions}. The overhead is mainly caused by the checks on synchronization instructions.

\subsection{Fault Injection Experiments}
\label{sec:fi_eval}

\begin{figure}[t]
\centering
\includegraphics[scale=0.7]{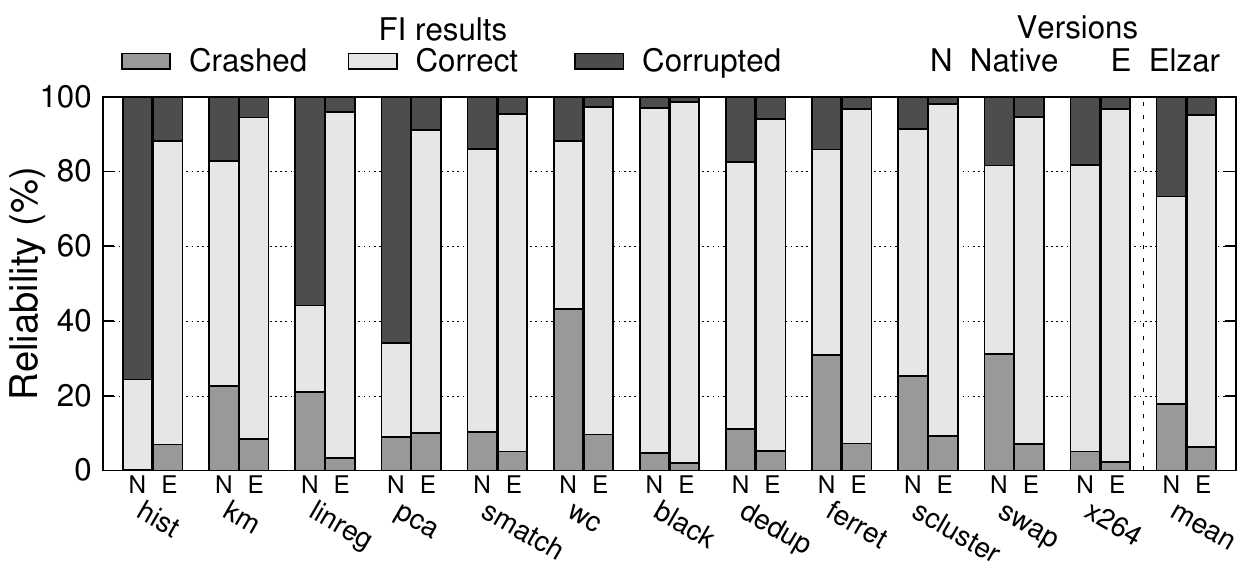}
\vspace{-5mm}
\caption{Reliability of \projectname{} (fault injections done on benchmarks with 2 threads).
}\label{fig:benches-fi}
\vspace{2mm}
\end{figure}
 \begin{figure}[t]
\centering
\includegraphics[scale=0.7]{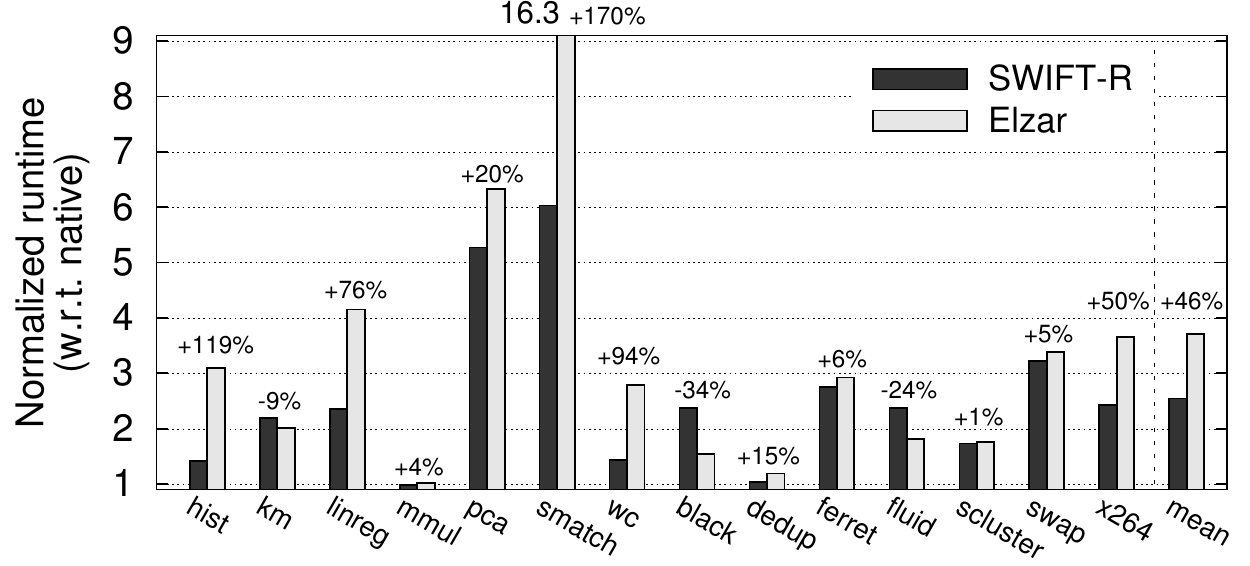}
\vspace{-5mm}
\caption{Performance comparison of \projectname{} and \swiftr{} (with 16 threads).
}\label{fig:benches-comparison}
\vspace{2mm}
\end{figure}
 
The results of the fault injection experiments are shown in \figref{benches-fi}. On average, \projectname{} reduces the SDC rate from $27\%$ to $5\%$ and the crash rate from $18\%$ to $6\%$.

\emph{Histogram} has the worst result with $12\%$ SDC. It highlights \projectname's window of vulnerability: address extractions before loads and stores.
If a fault occurs in the extracted address, it will be used to load a value from the wrong address, and this value will then be broadcast to all replicas. In other words, the fault will remain undetected and may lead to SDC (similarly, such a fault may lead to a segmentation fault and therefore to a system crash). Indeed, \tabref{benches-stats} tends to confirm this observation since \emph{histogram} has the highest number of memory accesses among all benchmarks. Similarly, \emph{blackscholes} has the least number of loads/stores and thus has only $1\%$ SDC.

\begin{figure*}[t]
\centering
\includegraphics[scale=0.7]{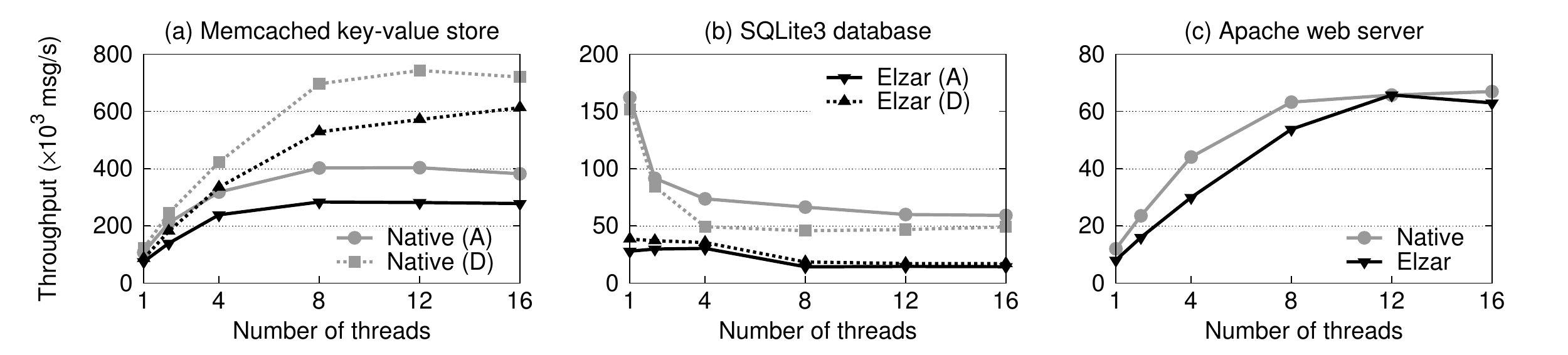}
\vspace{2mm}
\caption{Throughput of case studies: (a) Memcached key-value store, (b) SQLite3 database, and (c) Apache web server. Two extreme YCSB workloads are shown for Memcached and SQLite3: workload A ($50\%$ reads, $50\%$ writes, Zipf distribution) and workload D ($95\%$ reads, $5\%$ writes, latest distribution).
}\label{fig:casestudies-scalability}
\vspace{2mm}
\end{figure*}
 
\subsection{Comparison with Instruction Triplication}
\label{sec:comp_eval}

Lastly, we compare \projectname{} against a common \ilr{} approach based on triplication of instructions. More specifically, we compare \projectname{} against \swiftr{} \cite{SwiftR2007} as shown in \figref{benches-comparison}. We re-implemented \swiftr{} because its source code was not publicly available; we employed manual assembly inspection to make sure our implementation of \swiftr{} produces fast and correct code.

In general, \swiftr{} incurs lower overheads than \projectname{}, $2.5\times$ against $3.7\times$ on average. Interestingly, \projectname{} performs better in three benchmarks, namely \emph{kmeans}, \emph{blackscholes}, and \emph{fluidanimate}. To understand the differences between these approaches, we also report runtime statistics of \swiftr (\tabref{avxbenches-stats}).

We can draw two conclusions. First, \swiftr{} benefits from higher ILP, which is the key for its low performance overhead. As discussed before, \projectname{} takes a different stance and replicates not instructions but data; that is why it exhibits lower ILP but still performs on par with \swiftr{} in many cases.

Second, \swiftr{} significantly increases the number of instructions, which hampers its performance. \projectname{} has a smaller increase, proving our hypothesis that AVX-based \ilr{} leads to less code blow-up. For example, \projectname{} outperforms \swiftr{} on \emph{blackscholes} and \emph{fluidanimate} exactly for this reason: even though \swiftr's ILP is almost $2\times$ higher than \projectname, \swiftr produces $\sim2.5$--$3\times$ more instructions.

At the same time, \swiftr{} significantly outperforms \projectname{} in benchmarks that are dominated by memory accesses. In these cases, \projectname{} inserts a plethora of checks and wrappers, which results in a much higher number of instructions compared to \swiftr. This is exemplified by \emph{histogram}, \emph{string match}, and \emph{word count}.

 \section{Case Studies}
\label{sec:studies}

In this section, we report our experience on applying \projectname{} to three real-world applications: Memcached, SQLite3, and Apache.

\myparagraph{Memcached key-value store.} We evaluated Memcached v$1.4.24$ with all optimizations enabled, including atomic memory accesses. The evaluation was performed locally on the same Haswell machine used for other experiments, with $1$--$16$ cores dedicated to the Memcached server and all other cores to the YCSB clients \cite{YCSB2010} for generating workload. We opted to show the local performance of Memcached because the performance in a distributed environment is limited by the network and not by the CPU.

\figref{casestudies-scalability}a shows the throughput of native and \projectname{} versions of Memcached run with two extreme YCSB workloads: A ($50\%$ reads, $50\%$ writes, Zipf distribution) and D ($95\%$ reads, $5\%$ writes, latest distribution). We observe that \projectname{} scales on par with native, achieving up to $72\%$ of native throughput for workload A and up to $85\%$ for workload D. We also observed in our experiments that the latency of \projectname{} is $\sim25\%$ worse than native (not shown here). Such good results are explained partially by Memcached's poor memory locality, which amortizes the costs of \projectname.

\myparagraph{SQLite database.} We evaluated SQLite3 using an in-memory database and YCSB workloads, similar to Memcached. We should note that SQLite3 has a reverse scalability curve because it was designed to be thread-safe and \emph{not} concurrent. Therefore, SQLite3 exhibits worse throughput with higher numbers of threads.

The performance results are shown in \figref{casestudies-scalability}b. \projectname{} performs poorly, achieving only $20$--$30\%$ of the throughput of the native version. This overhead comes from the high number of locally near loads and stores, as well as function calls and function pointers. In all these cases, \projectname{} inserts additional checks and wrappers that significantly degrade performance.

\myparagraph{Apache web server.} We evaluated the Apache web server using its ``worker multi-processing module'' with a single running process and a varying number of worker threads. As a client, we used the classical \emph{ab} benchmark which repeatedly requests a static $1$MB web page.

\figref{casestudies-scalability}c shows the throughput with varying number of threads. \projectname{} performs very well, with an average throughput of $85\%$ compared to native. We attribute this good performance to the fact that Apache extensively uses third-party libraries that are not hardened by \projectname.

 \section{Discussion}
\label{sec:disc}

In this section, we highlight performance bottlenecks in the current AVX implementation and discuss the possible remedies.

\subsection{Performance Bottlenecks}
\label{sec:needforavx}

\myparagraph{Loads, stores, and branches.} Even not taking into account the overhead of checks, \projectname{} still performs $160\%$ worse than the native version (see \figref{benches-checks-impact}, ``all checks disabled''). This performance impact stems mainly from the three bottlenecks: loads,  stores, and branches.

To understand the impact of each of the three main bottlenecks, we created a set of microbenchmarks. Each microbenchmark has two versions: one with the regular instruction (e.g., regular load) and one with the AVX-based instruction (e.g., AVX-based load as shown in \figref{avxload}). In each microbenchmark, the instruction is replicated several times to saturate the CPU and wrapped in a loop to get execution time of at least 1 second. We wrote the microbenchmarks using volatile inline assembly to be sure that our instructions are not optimized away by the compiler; all tests were performed on our Intel Haswell machine.

The results of microbenchmarks are shown in \tabref{microbenches}. We conclude that adding \code{extract-broadcast} wrappers for AVX-based loads results in a $\sim2\times$ increase of load execution time. Similarly, adding \code{ptest} for AVX-based branches leads to an overhead of $\sim1.9\times$. Interestingly, AVX-based stores do not exhibit high overhead, which is explained by the fact that our Intel Haswell has only one port to process data stores and thus the store operation itself is a bottleneck even in the native version.

\begin{table}[t]
\renewcommand{\arraystretch}{0.8}
\centering

\begin{tabular}{l r r r}
 & \bfseries Loads & \bfseries Stores & \bfseries Branches \\
\hline
\hline
\noalign{\vskip 1pt}
average-case   & 1.96  & 1.00  & 1.86  \\
worst-case     & 2.06  & 1.14  & 1.89  \\
\hline
\end{tabular}

\caption{Normalized runtime of AVX-based versions of microbenchmarks w.r.t. native versions.}\label{tab:microbenches}
\vspace{2mm}
\end{table}
 
\myparagraph{Checks on loads and stores.} As can be seen from \figref{benches-checks-impact}, \projectname's checks on synchronization instructions contribute a significant amount of the overhead ($39\%$ on average). Specifically, checks on loads and stores account for most of the overhead because of the complicated sequence of check instructions (see \figref{avxcommoncheck}). At the same time, checks on branches add only $5\%$ overhead due to an efficient re-use of \code{ptest} already needed for branching itself (see \figref{avxbranchcheck}).

\myparagraph{Missing instructions.} Our Intel Haswell supports the AVX2 instruction set. Though AVX2 provides instructions for almost all operations, some classes of operations are missing. Two prominent examples are integer division and integer truncation. In the case of integer divisions, \projectname{} generates at least four regular division instructions and the corresponding wrappers to extract elements from the input YMM registers and insert elements in the output YMM register; with truncations, the situation is similar.
Clearly, emulating such missing instructions via a long sequence of available AVX instructions can lead to tremendous slowdowns.\footnote{One simple optimization would be to identify missing instructions and emit a sequence of only $3$ divisions/truncations. However, this solution still requires extracting elements and then combining them again. For our prototype, we had no need to implement such an optimization because these instructions are rare.}
For example, our microbenchmark for truncation exhibits overheads of $8\times$.

\subsection{Proposed AVX Instructions}
\label{sec:proposedavx}

\projectname{} could greatly benefit from a rather restricted set of new AVX instructions as proposed next. The instructions we propose are not \projectname-specific and other applications can find use for them. Moreover, some of them are already introduced in the AVX-512 instruction set which will be available in Intel's upcoming CPUs.

\myparagraph{Loads and stores (gathers and scatters).} As is clear from \figref{avxload}, regular load instructions are restricted in that they require an address operand specified in a general-purpose register (GPR). \projectname{} would need an instruction that can load the elements of an output YMM register from several addresses specified in the corresponding elements of an input YMM register.

\begin{figure}[t]
\centering
\includegraphics[scale=0.48]{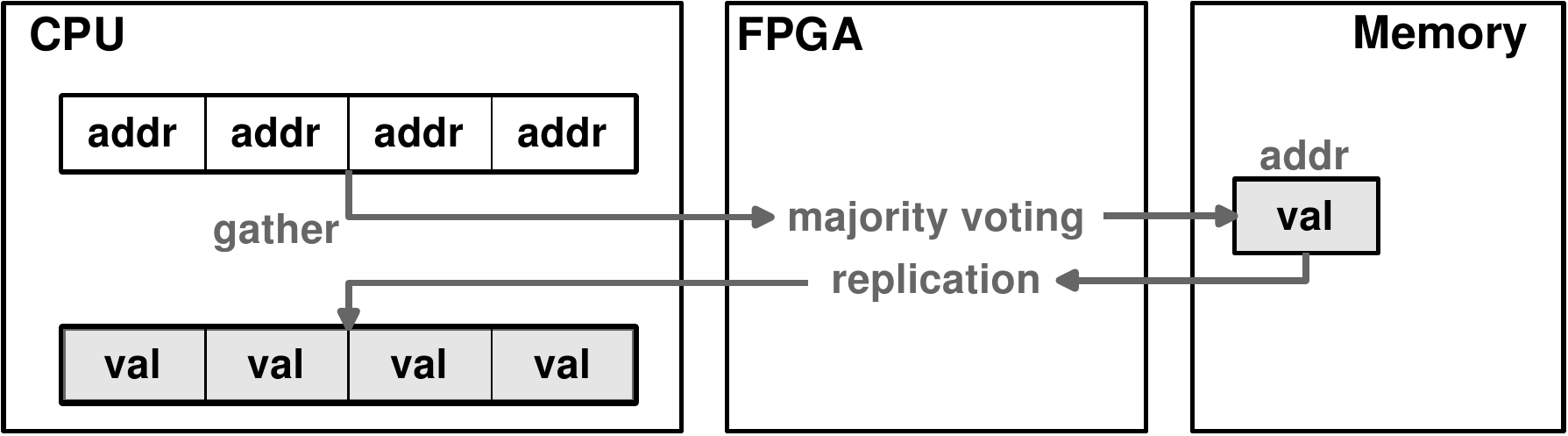}
\caption{Offloading checks to a FPGA accelerator via gather/scatter AVX instructions.}
\label{fig:fpgachecks}
\vspace{2mm}
\end{figure}

The current implementations of AVX already support a similar instruction called \code{gather} (\figref{fpgachecks}, left). Unfortunately, gather instructions still require a base address from a GPR and do not yet support all data types. Moreover, the current implementation is slower than a simple sequence of several loads \cite{Gather2014}. Nonetheless, we can expect that future AVX implementations will provide better support for gathers so that they can be successfully exploited in \projectname.
Interestingly, introducing gathers could also close a window of vulnerability discussed in \secref{fi_eval}.

A similar argument can be made regarding stores. AVX-512 introduces \code{scatter} instructions that can store elements from a YMM register based on the addresses in another YMM register. Thus, \projectname{} could advantageously substitute current implementations of stores with scatters.

\myparagraph{Comparisons affecting \flagsreg.} Currently, AVX exposes only one instruction, \code{ptest}, that can affect control flow by toggling the \flagsreg{} register. Accordingly, \projectname{} inserts an AVX-based comparison followed by a \code{ptest} to implement branching, as shown in \figref{avxbranch}. \tabref{microbenches} indicates that this additional operation leads to an overhead of almost $2\times$.

The only way to improve performance of branches is to re-implement the logic of the usual comparison instructions. In x86, a \code{cmp} instruction performs both the comparison \emph{and} the toggling of \flagsreg. We would propose a similar family of AVX-based comparisons which could output the result of comparison (\secref{simd_back}) \emph{and} set the corresponding flags in \flagsreg. Such improved comparisons could be also beneficial for vectorized applications that rely heavily on \code{ptest}.

\myparagraph{Checks on loads and stores.} Checks on loads and stores are implemented via an inefficient \code{shuffle-xor-ptest} sequence (see \figref{avxcommoncheck}). Having a single comparison instruction similar to the comparisons described above would greatly decrease the overheads of checks. Such an instruction would perform a pair-wise comparison of neighboring elements in a YMM register (so-called ``horizontal'' comparison) and toggle \flagsreg. Thus, a long sequence of instructions from \figref{avxcommoncheck} would be replaced by a single instruction.

The benefits of such an instruction for other applications than \projectname{} are unclear. Thus, in the next section we propose a more viable alternative involving an FPGA accelerator.

\myparagraph{Truncations, divisions, and others.} Curiously, a family of truncation operations (\code{vpmov}, \code{vcvt}) is already implemented in AVX-512. Integer division and modulo operations are quite rare and their absence is unlikely to lead to significant overheads; thus we believe these instructions are no candidates for future AVX implementations. We probably missed some other instructions that are not present in AVX, but we believe they are sufficiently uncommon to not provide much benefit for \projectname.

\subsection{Offloading Checks}
\label{sec:offloadchecks}

In order to decrease the overhead of checks, we can take advantage of the upcoming FPGA accelerators that will become part of CPUs \cite{XeonFPGA2015}. These FPGAs will be tightly coupled with the CPU and both will share the virtual memory of a process. As such, it will likely be possible to offload some functionality from the CPU to the FPGA.\footnote{As of December 2015, details on the Intel FPGA accelerators are not public and our speculations may prove wrong when the final products are released.}

We propose to offload the checks on loads and stores to the FPGA as follows (see \figref{fpgachecks}). For an \projectname-hardened program, all loads and stores are tunneled through the FPGA. The FPGA checks all copies of the address (for loads) and all copies of the value (for stores) and implements majority voting to mask possible faults. After that, the FPGA performs a load from a correct address or a store of a correct value. For loads, the FPGA also replicates the loaded value and sends it back to the CPU.

\subsection{Expected Overheads}

To summarize, our proposed set of changes in the underlying hardware is as follows: (1)~using AVX-based gathers/scatters for loads/stores, (2)~using AVX-based comparisons that can directly toggle \flagsreg, and (3)~offloading checks on loads/stores onto an FPGA.

To understand the synergistic effect of the proposed changes, we performed the following experiment.
First, we note that it is not possible to substitute AVX-based loads, stores, and branches with cheaper alternatives without disrupting the original flow of our benchmarks. Thus, we do a ``reverse'' comparison, i.e., instead of accelerating \projectname, we decelerate the native versions by adding dummy inline assembly around loads, stores, and branches. The assembly we add consists of instructions that \projectname{} uses as wrappers (see \secref{elzar}), e.g., we add dummy \code{extract} and \code{broadcast} for each load and a dummy \code{ptest} for each branch.\footnote{Adding dummy assembly can affect code generation and the CPU pipeline, but on average produces an adequately accurate estimation.}
Consequently, the overhead of \projectname{} with regard to this impaired native version serves as a rough estimate of \projectname{} overheads with our proposed changes.

\begin{figure}[t]
\centering
\includegraphics[scale=0.7]{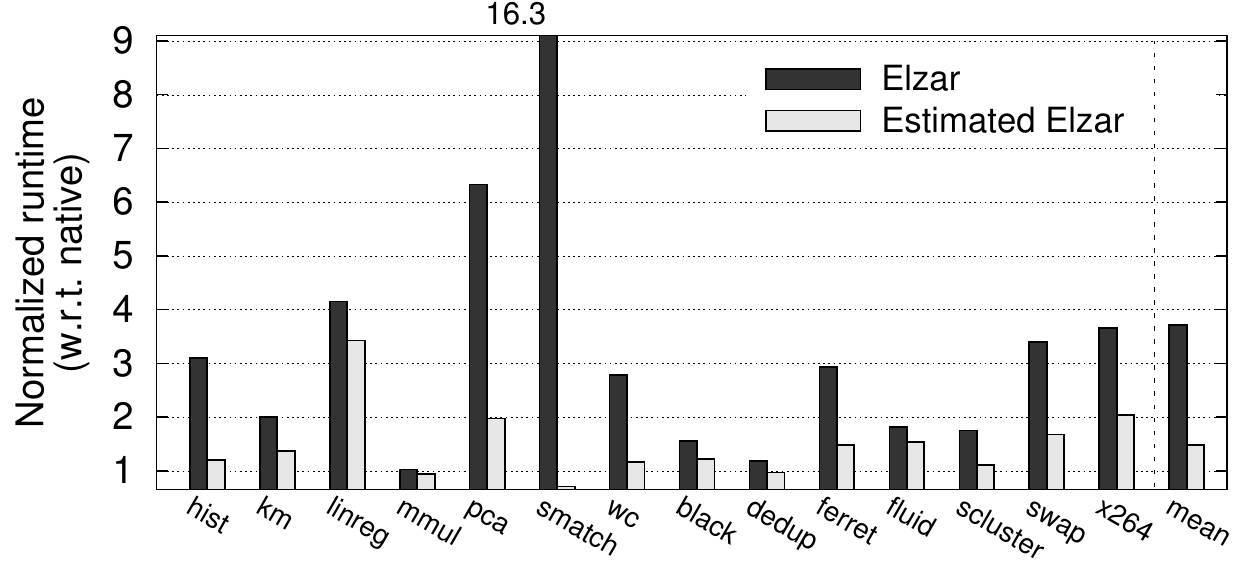}
\vspace{-6mm}
\caption{Estimation of performance overhead of \projectname{} with the proposed changes to AVX (with 16 threads).
}\label{fig:benches-slownative}
\vspace{2mm}
\end{figure}
 
The results of this experiment are shown in \figref{benches-slownative}. The average performance overhead is estimated to be $48\%$, i.e., an improvement of $150\%$ over current \projectname. Many benchmarks exhibit very low overhead of $10$--$20\%$. The case of \emph{string match} is peculiar, since it turns out to be faster than the native version in our experiment. Upon reading the disassembly, we found out that our dummy inline assembly in the ``decelerated'' native version prevented an optimization of function inlining: this led to a faster execution time of the \projectname version than the ``decelerated'' version. \section{Conclusion}

We presented \projectname, an AVX-based implementation of Instruction-Level Replication (\ilr).
\projectname{} achieves fault tolerance not by replicating instructions, but by replicating data inside AVX registers.
To our disappointment, we found out that AVX suffers from several limitations that lead to poor performance when used for \ilr.
The observed performance bottlenecks are primarily caused by the lack of suitable control flow and memory access instructions in the AVX instruction set, which necessitates the introduction of wrappers and ineffective checks for some types of instructions. We believe that these limitations can be overcome by simple extensions to the AVX instruction set.
We proposed improvements for the future generations of AVX that can lower the overheads of \projectname{} down to $\sim48\%$ according to our study.  

The shortened version of this report was published as a Practical Experience Report in the 46th Annual IEEE/IFIP International Conference on Dependable Systems and Networks (DSN'2016) \cite{elzar2016}.
 
\bibliographystyle{IEEEtran}
\bibliography{refs}

\begin{thebibliography}{10}
\providecommand{\url}[1]{#1}
\csname url@samestyle\endcsname
\providecommand{\newblock}{\relax}
\providecommand{\bibinfo}[2]{#2}
\providecommand{\BIBentrySTDinterwordspacing}{\spaceskip=0pt\relax}
\providecommand{\BIBentryALTinterwordstretchfactor}{4}
\providecommand{\BIBentryALTinterwordspacing}{\spaceskip=\fontdimen2\font plus
\BIBentryALTinterwordstretchfactor\fontdimen3\font minus
  \fontdimen4\font\relax}
\providecommand{\BIBforeignlanguage}[2]{{%
\expandafter\ifx\csname l@#1\endcsname\relax
\typeout{** WARNING: IEEEtran.bst: No hyphenation pattern has been}%
\typeout{** loaded for the language `#1'. Using the pattern for}%
\typeout{** the default language instead.}%
\else
\language=\csname l@#1\endcsname
\fi
#2}}
\providecommand{\BIBdecl}{\relax}
\BIBdecl

\bibitem{CPUFailures2005}
G.~Saggese, N.~Wang, Z.~Kalbarczyk, S.~Patel, and R.~Iyer, ``An experimental
  study of soft errors in microprocessors,'' in \emph{Micro}, 2005.

\bibitem{Borkar2005}
S.~Borkar, ``Designing reliable systems from unreliable components: The
  challenges of transistor variability and degradation,'' in \emph{Micro},
  2005.

\bibitem{Henkel2013}
J.~Henkel, L.~Bauer, N.~Dutt, P.~Gupta, S.~Nassif, M.~Shafique, M.~Tahoori, and
  N.~Wehn, ``Reliable on-chip systems in the nano-era: Lessons learnt and
  future trends,'' in \emph{DAC}, 2013.

\bibitem{DarkSilicon2014}
M.~Shafique, S.~Garg, J.~Henkel, and D.~Marculescu, ``The {EDA} challenges in
  the {Dark Silicon} era: Temperature, reliability, and variability
  perspectives,'' in \emph{DAC}, 2014.

\bibitem{schroeder2010large}
B.~Schroeder, G.~Gibson \emph{et~al.}, ``A large-scale study of failures in
  high-performance computing systems,'' in \emph{TDSC}, 2010.

\bibitem{CyclesCells2011}
E.~B. Nightingale, J.~R. Douceur, and V.~Orgovan, ``Cycles, cells and platters:
  An empirical analysis of hardware failures on a million consumer {PC}s,'' in
  \emph{EuroSys}, 2011.

\bibitem{BugsInCloud2014}
H.~S. Gunawi, M.~Hao, T.~Leesatapornwongsa, T.~Patana-anake, T.~Do,
  J.~Adityatama, K.~J. Eliazar, A.~Laksono, J.~F. Lukman, V.~Martin, and A.~D.
  Satria, ``What bugs live in the cloud? a study of 3000+ issues in cloud
  systems,'' in \emph{SoCC}, 2014.

\bibitem{AmazonS32008}
``Amazon {S3} availability event,''
  \url{http://status.aws.amazon.com/s3-20080720.html}, accessed: Dec, 2015.

\bibitem{AmazonLoadBalancer2008}
``New defective {S3} load balancer corrupts relayed messages,''
  \url{https://forums.aws.amazon.com/thread.jspa?threadID=22709}, accessed:
  Oct, 2015.

\bibitem{Mesa2014}
{A. Gupta et al.}, ``Mesa: Geo-replicated, near real-time, scalable data
  warehousing,'' in \emph{VLDB}, 2014.

\bibitem{BFTSceptics2009}
Y.~J. Song, F.~P. Junqueira, and B.~Reed, ``{BFT} for the skeptics,'' in
  \emph{BFTW3}, 2009.

\bibitem{Bhatotia2010}
P.~Bhatotia, A.~Wieder, R.~Rodrigues, F.~Junqueira, and B.~Reed, ``Reliable
  data-center scale computations,'' in \emph{Proceedings of the 4th
  International Workshop on Large Scale Distributed Systems and Middleware
  (LADIS)}, 2010.

\bibitem{EDDI2002}
N.~Oh, P.~Shirvani, and E.~McCluskey, ``Error detection by duplicated
  instructions in super-scalar processors,'' in \emph{Transactions on
  Reliability}, 2002.

\bibitem{Swift2005}
G.~A. Reis, J.~Chang, N.~Vachharajani, R.~Rangan, and D.~I. August, ``{SWIFT}:
  Software implemented fault tolerance,'' in \emph{CGO}, 2005.

\bibitem{haft2016}
D.~Kuvaiskii, R.~Faqeh, P.~Bhatotia, P.~Felber, and C.~Fetzer, ``{HAFT:
  Hardware-Assisted Fault Tolerance},'' in \emph{Eurosys}, 2016.

\bibitem{SwiftR2007}
G.~A. Reis, J.~Chang, and D.~I. August, ``Automatic instruction-level
  software-only recovery,'' in \emph{Micro}, 2007.

\bibitem{Phoenix2007}
C.~Ranger, R.~Raghuraman, A.~Penmetsa, G.~Bradski, and C.~Kozyrakis,
  ``Evaluating {MapReduce} for multi-core and multiprocessor systems,'' in
  \emph{HPCA}, 2007.

\bibitem{Parsec2009}
C.~Bienia and K.~Li, ``{PARSEC 2.0}: A new benchmark suite for
  chip-multiprocessors,'' in \emph{MoBS}, 2009.

\bibitem{HWRMT2002}
S.~Mukherjee, M.~Kontz, and S.~Reinhardt, ``Detailed design and evaluation of
  redundant multi-threading alternatives,'' in \emph{ISCA}, 2002.

\bibitem{DAFT2010}
Y.~Zhang, J.~W. Lee, N.~P. Johnson, and D.~I. August, ``{DAFT}: Decoupled
  acyclic fault tolerance,'' in \emph{PACT}, 2010.

\bibitem{PLR2007}
A.~Shye, T.~Moseley, V.~Reddi, J.~Blomstedt, and D.~Connors, ``Using
  process-level redundancy to exploit multiple cores for transient fault
  tolerance,'' in \emph{DSN}, 2007.

\bibitem{RAFT2012}
Y.~Zhang, S.~Ghosh, J.~Huang, J.~W. Lee, S.~A. Mahlke, and D.~I. August,
  ``Runtime asynchronous fault tolerance via speculation,'' in \emph{CGO},
  2012.

\bibitem{RomainMT2014}
B.~D\"{o}bel and H.~H\"{a}rtig, ``Can we put concurrency back into redundant
  multithreading?'' in \emph{EMSOFT}, 2014.

\bibitem{ESoftCheck2009}
J.~Yu, M.~J. Garzaran, and M.~Snir, ``{ESoftCheck}: Removal of non-vital checks
  for fault tolerance,'' in \emph{CGO}, 2009.

\bibitem{Shoestring2010}
S.~Feng, S.~Gupta, A.~Ansari, and S.~Mahlke, ``Shoestring: Probabilistic soft
  error reliability on the cheap,'' in \emph{ASPLOS}, 2010.

\bibitem{Chen2015}
Z.~Chen, R.~Inagaki, A.~Nicolau, and A.~Veidenbaum, ``Software fault tolerance
  for {FPUs} via vectorization,'' in \emph{SAMOS}, 2015.

\bibitem{TMR1962}
R.~Lyons and W.~Vanderkulk, ``The use of triple-modular redundancy to improve
  computer reliability,'' in \emph{IBM Journal of Research and Development},
  1962.

\bibitem{LLVM2004}
C.~Lattner and V.~Adve, ``{LLVM}: A compilation framework for lifelong program
  analysis and transformation,'' in \emph{CGO}, 2004.

\bibitem{ParsecPerf2009}
M.~Bhadauria, V.~M. Weaver, and S.~A. McKee, ``Understanding {PARSEC}
  performance on contemporary {CMP}s,'' in \emph{IISWC}, 2009.

\bibitem{YCSB2010}
B.~F. Cooper, A.~Silberstein, E.~Tam, R.~Ramakrishnan, and R.~Sears,
  ``Benchmarking cloud serving systems with {YCSB},'' in \emph{SoCC}, 2010.

\bibitem{Gather2014}
J.~Hofmann, J.~Treibig, G.~Hager, and G.~Wellein, ``Comparing the performance
  of different x86 {SIMD} instruction sets for a medical imaging application on
  modern multi- and manycore chips,'' in \emph{WPMVP}, 2014.

\bibitem{XeonFPGA2015}
P.~Gupta, ``{Xeon+FPGA} platform for the data center,''
  \url{http://www.ece.cmu.edu/~calcm/carl/lib/exe/fetch.php?media=carl15-gupta.pdf},
  accessed: Dec, 2015.

\bibitem{elzar2016}
D.~Kuvaiskii, O.~Oleksenko, P.~Bhatotia, P.~Felber, and C.~Fetzer, ``{Elzar:
  Triple Modular Redundancy using Intel AVX},'' in \emph{DSN}, 2016.

\end{thebibliography}

\end{document}